\documentclass[aps,twocolumn]{revtex4-2}
\usepackage{amsmath}
\usepackage{graphicx}
\usepackage{enumitem}
\usepackage[top = 1. in,bottom = 1. in, left = 1 in, right=1 in]{geometry}
\usepackage[colorlinks=true, linkcolor=blue, bookmarks=true]{hyperref}

\allowdisplaybreaks

\makeatletter
\renewcommand\section{\@startsection {section}{1}{\z@}   {-3.5ex \@plus -1ex \@minus -.2ex}   {2.3ex \@plus.2ex}   {\normalfont\large\bfseries}}
\renewcommand\subsection{\@startsection{subsection}{2}{\z@}   {-3.25ex\@plus -1ex \@minus -.2ex}   {1.5ex \@plus .2ex}   {\normalfont\bfseries}}
\makeatother

\newcommand{\de}{\delta}
\newcommand{\beq}{\begin{equation}}
\newcommand{\eeq}{\end{equation}}
\newcommand{\ber}{\begin{array}}
\newcommand{\eer}{\end{array}}

\newcommand{\ena}{\end{eqnarray}}
\newcommand{\beqa}{\begin{eqnarray}}
\newcommand{\eeqa}{\end{eqnarray}}
\newcommand{\bea}{\begin{eqnarray}}
\newcommand{\eea}{\end{eqnarray}}

\begin{document}

\title{Fermi-Pasta-Ulam phenomena and persistent breathers in the harmonic
trap}
\author{Anxo Biasi$^{1}$, Oleg Evnin$^{2,3}$, Boris A. Malomed$^{4,5}$%
\vspace{2mm}}
\affiliation{ $^{1}$Institute of Theoretical Physics, Jagiellonian University, Krakow 30-348, Poland\\
	$^{2}$Department~of~Physics,~Faculty~of~Science,~Chulalongkorn~University,~Bangkok 10330,~Thailand\\
$^{3}$Theoretische Natuurkunde, Vrije Universiteit Brussel and International Solvay Institutes, Brussels 1050, Belgium\\
$^{4}$Department of Physical Electronics, School of Electrical Engineering,
Tel Aviv University, Tel Aviv 69978, Israel\\
$^{5}$Instituto de Alta Investigaci\'{o}n, Universidad de Tarapac\'{a}, Casilla 7D, Arica, Chile}

\begin{abstract}
We consider the long-term weakly nonlinear evolution governed by the
two-dimensional nonlinear Schr\"{o}dinger (NLS) equation with an isotropic
harmonic oscillator potential. The dynamics in this regime is dominated by
resonant interactions between quartets of linear normal modes, accurately
captured by the corresponding resonant approximation. Within this approximation, 
we identify Fermi-Pasta-Ulam-like recurrence phenomena, whereby the
normal-mode spectrum passes in close proximity of the initial configuration,
and two-mode states with time-independent mode amplitude spectra that
translate into long-lived breathers of the original NLS equation. We comment
on possible implications of these findings for nonlinear optics and
matter-wave dynamics in Bose-Einstein condensates.
\end{abstract}

\maketitle

\section{Introduction}

Nonlinear Schr\"{o}dinger (NLS) equations of the form
\begin{equation}
i\partial _{t}\Psi =\left( -\frac{1}{2}\nabla ^{2}+V(\mathbf{x})\right) \Psi
+g|\Psi |^{2}\Psi ,  \label{eq:GPgen}
\end{equation}%
with real potentials $V(\mathbf{x})$ and cubic nonlinearity strength $g$
provide a universal framework for modeling a wealth of physical phenomena in
weakly nonlinear dispersive media, including the dynamics of matter waves in
Bose-Einstein Condensates (BECs), propagation of optical signals in
dielectric media, Langmuir excitations in plasmas, surface waves on deep
water, etc \cite{sulem2,Fibich}. For appropriate choices of $V(x)$ and
especially in lower spatial dimensions, these equations exhibit highly
organized dynamical phenomena, some of which we address in this work.

The best-studied form of dynamics in NLS equations occurs in the 1D setting
with $V(x)=0$, which exhibits integrability with its celebrated
manifestations in the form of exact multi-soliton \cite%
{1dtextbook,Ablowitz-Segur} and breather \cite{SY} solutions. The
integrability is, of course, a very fragile property. In particular, it is
destroyed by nonvanishing potentials $V(x)$ in the 1D version of Eq. (\ref%
{eq:GPgen}). Precise consequences of the integrability breaking
significantly depend on the shape of the potential \cite{RMP}.

It has been emphasized in \cite{quasiint2} that the harmonic-oscillator (HO)
potential is very special in this regard. While no form of integrability is
known to be valid for the 1D NLS equation with the HO potential, the
dynamics of this system is very far from the ergodic form generic to
non-integrable Hamiltonian systems. In particular, systematic simulations
reveal that the NLS equation with a self-repulsive nonlinearity, $g>0$ in
Eq. (\ref{eq:GPgen}), displays a quasi-discrete dynamical power spectrum,
unlike the continuous spectra typical for non-integrable dynamics \cite%
{quasiint2}. As an empiric effect, the absence of \textquotedblleft
turbulence" in simulations of this model was observed in earlier works
\cite{no-chaos1,no-chaos2}. This phenomenon, referred to as
\textquotedblleft quasi-integrability" \cite{quasiint2}, does not occur with
other forms of trapping potentials, e.g., the potential box, which readily
give rise to the usual ergodic dynamics and continuous power spectra \cite%
{chaos1,chaos2,quasiint2}.

The special role of the HO potential is retained in two spatial dimensions
(2D), in which case some regular, highly organized motions are observed, in
contrast to the egodicity, which, as mentioned above, one may expect in a
generic non-integrable system. These motions include periodic splitting and
recombination of unstable vortices \cite{quasiint1} as well as a range of
breather solutions \cite{tri}. In the Thomas-Fermi limit, the breathers of \cite{tri}
include strikingly simple circular and triangular configurations with sharp
boundaries. While the emergence of these breathers has been given an
analytic explanation in \cite{tri1,tri2,Zhou}, and similarly
case-by-case analytic understanding has been developed for some regular
motions in other cases, as shown below, we are not aware of any overarching
mathematical structure (as in the case of integrable equations) that would
underlie such regular dynamics. Physically, the persistence of regular
dynamics and the absence of ergodicity are related to various obstructions
to thermalization of low-dimensional interacting multi-boson systems, which
occur in a variety of physically relevant settings \cite%
{no-chaos2,atomtherm,atomtherm1,atomtherm2}.

In the present work, we focus on Eq. (\ref{eq:GPgen}) in 2D with the
isotropic HO potential in the weakly nonlinear regime. It is well known as
the mean-field Gross-Pitaevskii (GP)\ equation for BECs in pancake-shaped
ultracold atomic gases, strongly confined by an external field in the third
direction \cite{Pit,BDZ}. The same equation with $t$ replaced by the
propagation distance $z$ governs the transmission of light beams through a
bulk waveguide with transverse coordinates $\left( x,y\right) $, where the
HO potential represents the guiding profile of the local refractive index
\cite{Agrawal}. Similar equations have been considered from theoretical and
mathematical perspectives in higher spatial dimensions, where they exhibit
noteworthy dynamical phenomena \cite{BMP,PPF}.

Weakly nonlinear dynamics in HO potentials is quite peculiar because the
perfectly resonant frequency spectrum of the corresponding linearized
problem (the usual equidistant spectrum of the quantum HO) leads to a
dramatic enhancement of weak nonlinearities. Generically, weakly nonlinear
evolution can be thought of as quasi-linear evolution, in which the
amplitudes and phases of the normal modes are not constant, but undergo slow
modulations under the action of the weak nonlinearity. For highly resonant
spectra of the linearized-normal-mode frequencies, such as the HO spectrum,
the slow modulations may accumulate to effects of order $1$ for small $g$ in
Eq. (\ref{eq:GPgen}), on time scales $\sim 1/g$. Such large effects of small
nonlinearities on long temporal scales are effectively captured by
simplified \textit{resonant systems}, whose dynamics is the main subject of
this work. Resonant systems are widely used for
weakly nonlinear analysis of highly resonant PDEs, and are known, 
in various branches of research, as the multi-scale analysis, time-averaging, or effective
equations \cite{murdock,KM}.  For the NLS/GP
equation with the HO potential in 2D, rigorous mathematical proofs have been
developed \cite{GHT} for the accuracy of the resonant system as an
approximation to the original PDE in the relevant weakly nonlinear regime;
see also Ref. \cite{fennell} for a similar treatment in 1D. Note that
restrictions to resonant interactions between the modes are also essential to
the wave turbulence theory \cite{Nazarenko}, though in that setting the phases
of normal modes are treated as random variables, while the resonant
approximation as considered here is fully deterministic.

Within the resonant approximation, the dynamics of the 2D NLS/GP
equation with the isotropic HO potential has been previously analyzed in
\cite{BBCE,GGT,BBCE2}, where its evolution was found to display a
variety of regular dynamical patterns, time-periodic and stationary.
They take the form of precessing vortices \cite{BBCE}, oscillating rings
\cite{BBCE2}, as well as revolving and precessing vortex arrays \cite{BBCE2}%
. General results on positions of vortices for configurations that are
stationary within the resonant approximation have been presented in \cite%
{GGT}. These analytic results rely on very special properties of mode
couplings for the NLS/GP equation with the 2D HO potential, and some general
mathematical structures underlying these simple behaviors have been
uncovered in \cite{AO,BBE2,E}. The corresponding quantum many-body
problems, considered outside the GP-based mean-field approximation, likewise
display pronounced regular features \cite{CCEK,MO,RPP} (see also the review \cite%
{RMP}).

Our purpose in this work is to present new dynamical regimes for the 2D
NLS/GP equation with the isotropic HO potential, approximated by the
corresponding resonant system, beyond those reported in \cite{BBCE,GGT,BBCE2}%
. It is natural to consider the weakly nonlinear evolution in terms of the
slow energy transfer between linearized normal modes. This perspective
suggests the question whether Fermi-Pasta-Ulam (FPU) phenomena \cite%
{FPUorg,FPUrev} may occur in our setting for some initial data. The
notion of FPU dynamics goes back to the classic paper \cite{FPUorg} where it
was observed that the distribution of energy among the normal modes of weakly
nonlinear oscillator chains returns, in some situations, to close
proximity of the initial configuration. The energy thus fails to effectively
redistribute among all available degrees of freedom, as would be suggested by
ergodicity (thermalization). Here, we address the question whether similar
phenomena occur in the evolution of the mode spectrum of the NLS/GP
resonant system.

There are a few reasons why one may expect FPU phenomena to occur in the
present case. First, in Refs. \cite{BBCE,BBCE2}, perfect (rather than
approximate) returns of the amplitude spectrum to the initial state have
been observed for some very specific initial data in the framework of the
resonant approximation that we consider here. Second, FPU-like approximate
returns have been reported \cite{FPU,BCE} for relativistic analogs of the
NLS/GP equations with the HO potentials (those systems, defined in the
anti-de Sitter spacetime, reduce to the NLS equation in the nonrelativistic
limit \cite{CEL,BEF,resrev}, hence they have essentially identical
normal-mode spectra, and their resonant approximation \cite{CEV1,CEV2,BMR,CF}
is structured identically to that for the NLS/GP problems, as originally
pointed out in Ref. \cite{BMP}). Third, FPU returns have been observed in a
related setup in Ref. \cite{AngelFPUT}, albeit in the absence of the HO trap
and with two different nonlinear terms included in the equation.

We thus set the goal of identifying the FPU dynamics within the resonant
approximation for the 2D HO-trapped NLS/GP equation, and reporting initial
configurations that result in this dynamics. While looking for FPU returns,
we additionally discover infinite families of two-mode initial data that lead to no
energy transfer at all within the resonant approximation, due to the
vanishing of specific four-mode couplings. These configurations produce
long-lived breathers in terms of the underlying NLS equation, which are
of interest in their own right. Before proceeding with the
presentation of the technical results, we introduce in section~\ref%
{sec:setup} the setup for analyzing the weakly nonlinear resonant NLS/GP\
dynamics, and then review, in section~\ref{sec:trunc}, how the dynamics can
be consistently restricted to smaller sets of modes in which the target
phenomena are observed (if only modes from one of these sets are excited in
the initial state, subsequent evolution does not excite any modes outside
the set). In sections~\ref{sec:fpu} and \ref{sec:breath}, we present our
findings for the FPU returns and two-mode breathers, respectively. The paper
is concluded by section \ref{sec:conclusion}, which includes a brief
discussion of possible applications.


\section{The weakly nonlinear resonant evolution}

\label{sec:setup}

We consider the 2D NLS/GP equation with the isotropic HO potential, written
in the scaled form:
\begin{equation}
i\partial _{t}\Psi =\frac{1}{2}\left( -\nabla ^{2}+r^{2}\right) \Psi +g|\Psi
|^{2}\Psi ,  \label{eq:GP}
\end{equation}%
where $r$ is the radial coordinate, cf. Eq. (\ref{eq:GPgen}). We focus on
the weak-coupling regime with $|g|\ll 1$, and consider the evolution on long
time scales $\sim 1/|g|$. The weakly nonlinear evolution amounts to slow
modulations of amplitudes and phases of the linearized normal modes. Because
of the resonant nature of the spectrum of the ordinary linear Schr\"{o}%
dinger equation for the HO, arbitrarily small nonlinearities may generate
effects of order $1$ on such time scales. Leading effects of this sort are
captured by the \textit{resonant approximation} that we construct below. In
this context, opposite signs of the coupling $g$ lead to essentially
identical modulation patterns (up to time reversal). Of course, the sign
of $g$ leads to drastic differences at finite values of $|g|$, such as the
occurrence of critical collapse under the action of the focusing term
with $g<0$ \cite{sulem2,Fibich}, but in the weakly nonlinear regime that we
address here, such differences appear on time scales $t\gg
1/|g| $, which are outside of the scope of our analysis.

The 2D NLS/GP equation (\ref{eq:GP}) with the HO trap and cubic nonlinearity
is rather special as it features a symmetry enhancement in comparison to
other dimensions, manifested, in particular, by the presence of the
Pitaevskii-Rosch breathing mode \cite{breathing,qbreathing}. In close
relation to this breathing mode is the lens transform \cite{Talanov,Papa},
also known as the \textquotedblleft pseudoconformal
compactification\textquotedblright\ \cite{Tao09}, that allows one to map
into each other the evolution governed by the 2D equation (\ref{eq:GP}), and the
same equation without the potential term (the mapping relates infinite and
finite time intervals for the two equations involved). In particular, this mapping
has been recently employed in \cite{tri2}.

\begin{figure}[t]
\begin{center}
\includegraphics[width=\columnwidth]{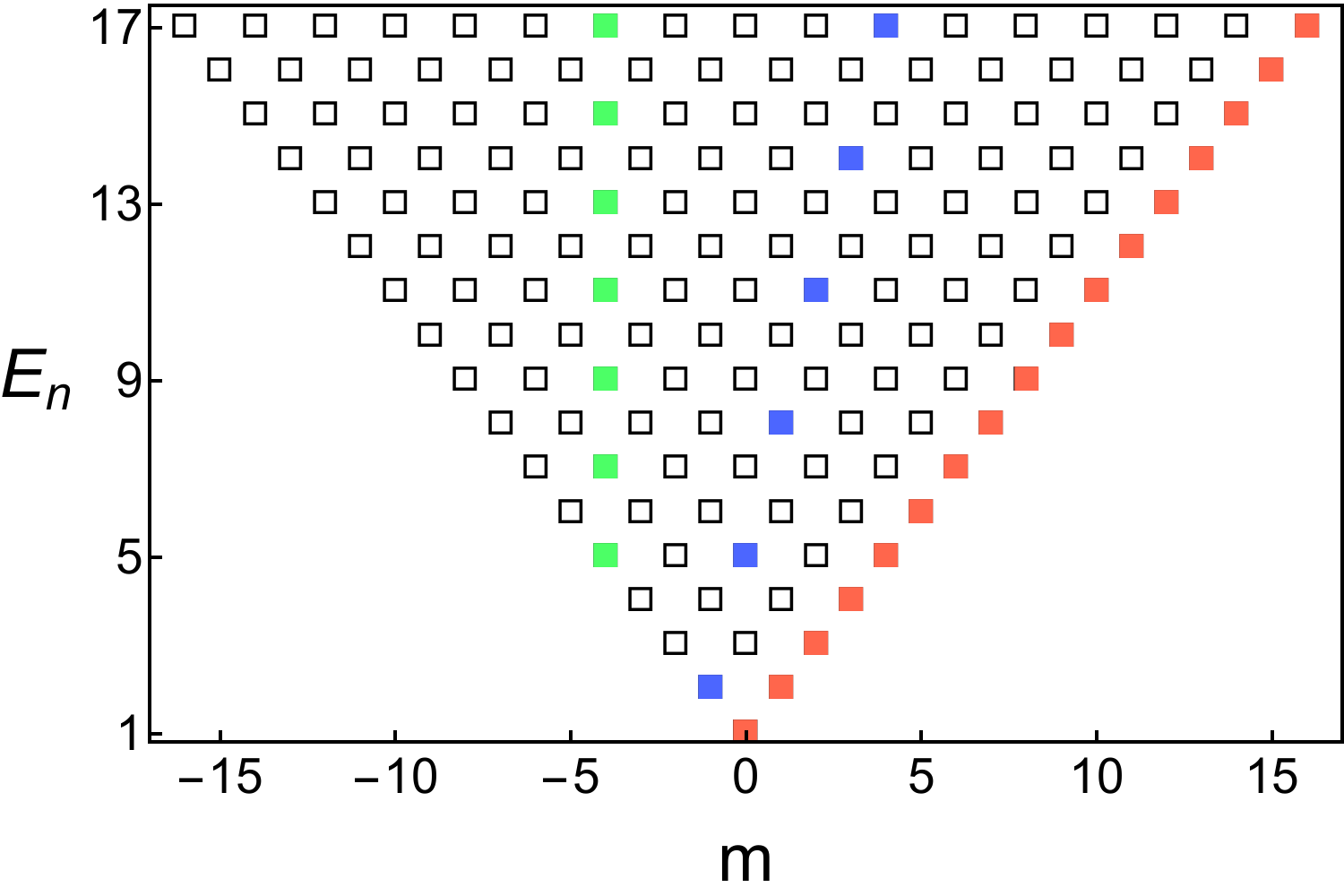}\vspace{-5mm}
\end{center}
\par
\vspace{-2mm}
\caption{Energies (\protect\ref{En}) corresponding to normal modes (\protect
\ref{eq:linear_modes}) labeled by quantum numbers $n$ and $m$. A few
possible restrictions to smaller sets of modes, addressed within the
resonant approximation in section~\protect\ref{sec:trunc}, are highlighted
by colors. The rightmost highlighted diagonal line is the lowest
Landau-level truncation which was used in Ref. \protect\cite{BBCE,GGT}. The
vertical highlighted line represents the fixed-angular-momentum truncations
employed in Ref. \protect\cite{BBCE2}. The remaining highlighted line
represents a generic restriction to a \textquotedblleft one-dimensional"
subset of modes that play a central role here, in sections~\protect\ref%
{sec:fpu} and \protect\ref{sec:breath}.}
\label{fig:Diagram_nm}
\end{figure}

Before proceeding with the weakly nonlinear analysis, we write down the general
solution of the linearized problem ($g=0$), composed of the HO eigenstates
\cite{dahl}:
\begin{equation}
\Psi _{nm}^{\mbox{\tiny norm.}}=\sqrt{\frac{(\frac{1}{2}(n-|m|))!}{(\frac{1}{%
2}(n+|m|))!}}\frac{r^{|m|}}{\sqrt{\pi }}L_{\frac{n-|m|}{2}%
}^{|m|}(r^{2})e^{-r^{2}/2}e^{im\phi }.  \label{psipos}
\end{equation}%
Here, $(r,\phi )$ are the polar coordinates, $L_{n}^{\alpha }(x)$ are the
generalized Laguerre polynomials, $n=0,1,2,...$ labels the energy level
\begin{equation}
E_{n}=n+1,  \label{En}
\end{equation}%
and $m\in \{-n,-n+2,...,n-2,n\}$ labels the angular momentum. To optimize
the subsequent analysis, we introduce an extra sign factor in the definition
of the eigenfunctions as follows:
\begin{equation}
\Psi _{nm}=(-1)^{\frac{1}{2}(m-|m|)}\Psi _{nm}^{\mbox{\tiny norm.}}.
\end{equation}%
This factor brings our set of the linearized normal modes in accord with
Ref. \cite{GHT}. Using the identity
\begin{align*}
& \sqrt{\frac{((n-|m|)/2)!}{((n+|m|)/2)!}}r^{|m|}L_{\frac{n-|m|}{2}%
}^{|m|}(r^{2})e^{-r^{2}/2} \\
& =(-1)^{\frac{1}{2}(m-|m|)}\sqrt{\frac{((n-m)/2)!}{((n+m)/2)!}}r^{m}L_{%
\frac{n-m}{2}}^{m}(r^{2})e^{-r^{2}/2},
\end{align*}%
one obtains the following expression:
\begin{equation}
\Psi _{nm}=\sqrt{\frac{(\frac{1}{2}(n-m))!}{(\frac{1}{2}(n+m))!}}\frac{r^{m}%
}{\sqrt{\pi }}L_{\frac{n-m}{2}}^{m}(r^{2})e^{-r^{2}/2}e^{im\phi },
\label{eq:linear_modes}
\end{equation}%
which is the basis that will be employed below. The tower of linearized
normal modes is visualized in Fig.~\ref{fig:Diagram_nm}.

One could try to perturbatively improve, as a power series in $g$, the
linearized solutions of Eq. (\ref{eq:GP}),
\begin{equation}
\Psi _{\mbox{\tiny linear}}(t,r,\phi )=\sum_{n,m}A_{nm}e^{-iE_{n}t}\Psi
_{nm}(r,\phi ),  \label{eq:Psi_sum_linear}
\end{equation}%
where $\Psi _{nm}$ are the normal modes given by Eq. (\ref{eq:linear_modes})
and $A_{nm}$ are constant complex amplitudes. This approach is known to
fail, however, due to the apprearance of \textit{secular terms} that grow in
time and lead to breakdown of the perturbative expansion on time scales $%
t\sim 1/|g|$. An appropriate alternative that correctly captures the
dynamics on the relevant time scales is provided by the \textit{resonant
approximation}. To develop it, we start by decomposing exact solutions to
Eq. (\ref{eq:GP}) in terms of the linearized modes,%
\begin{equation}
\Psi (t,r,\phi )=\sum_{n,m}\alpha _{nm}(t)\,e^{-iE_{n}t}\Psi _{nm}(r,\phi ),
\label{eq:Psi_sum_linear_modes}
\end{equation}%
where $\alpha _{nm}$ are complex-valued functions of time. Plugging this
decomposition in Eq. (\ref{eq:GP}) and projecting onto $\Psi _{nm}$, one
obtains an infinite system of ordinary differential equations,
\begin{equation}
i\frac{d\alpha _{nm}}{dt}=g%
\sum_{n_{i}m_{i}}C_{nn_{1}n_{2}n_{3}}^{mm_{1}m_{2}m_{3}}\bar{\alpha}%
_{n_{1}m_{1}}\alpha _{n_{2}m_{2}}\alpha _{n_{3}m_{3}}e^{-iEt},
\label{eq:Resonant_Eq_v0}
\end{equation}%
where the bar stands for complex conjugation,
\begin{equation}
E\equiv E_{n}+E_{n_{1}}-E_{n_{2}}-E_{n_{3}},
\end{equation}%
and the \textit{interaction coefficients}, or re\-so\-nant mode couplings,
are defined by%
\begin{equation}
C_{nn_{1}n_{2}n_{3}}^{mm_{1}m_{2}m_{3}}\equiv \int_{0}^{\infty }\hspace{-3mm}%
rdr\int_{0}^{2\pi }\hspace{-3mm}d\phi ~\bar{\Psi}_{nm}\bar{\Psi}%
_{n_{1}m_{1}}\Psi _{n_{2}m_{2}}\Psi _{n_{3}m_{3}}.  \label{eq:C_nnnn_mmmm}
\end{equation}%
As the $\phi $-dependence of $\Psi _{nm}$ is given by $e^{im\phi }$, the $%
\phi $ integration nullifies $C$, unless
\begin{equation}
m+m_{1}=m_{2}+m_{3},
\end{equation}%
which represents the angular momentum conservation.

Note that we did not make use of the weakly nonlinear limit yet, and the
expressions up to this point are correct for finite values of $g$. On the
other hand, when $g$ is small the dynamics described by Eq. (\ref%
{eq:Resonant_Eq_v0}) acquires a conspicuous two-scale structure. Namely,
most terms on the right-hand side come with oscillatory factors $e^{-iEt}$
that vary on time scales $t\sim 1$, while $\alpha _{nm}$ have time
derivatives $\sim g$, thus varying slowly, on scales $t\sim 1/g$. It is
natural to expect that the effect of the oscillatory terms averages out and
may be neglected (a more precise statement is given below), while
significant contributions to the evolution of $\alpha _{nm}$ are produced
only by those terms on the right-hand side of Eq. (\ref{eq:Resonant_Eq_v0})
that do not oscillate on time scales $\sim 1$, which is precisely the terms
with $E=0$, or
\begin{equation}
n+n_{1}=n_{2}+n_{3}.  \label{eq:resonant_channels}
\end{equation}%
The \textit{resonant approximation}, so named after resonance condition (\ref%
{eq:resonant_channels}), is defined by keeping only such non-oscillatory
terms. Under this approximation, which is the main focus of our study, and
introducing the \textit{slow time} $\tau =gt$ (with overdots denoting $\tau $%
-derivatives from now on), one obtains the following \textit{resonant system}
of the 2D NLS/GP equation with the HO potential:
\begin{equation}
i\dot{\alpha}_{nm}=\hspace{-5mm}\underset{m+m_{1}=m_{2}+m_{2}}{%
\sum_{n+n_{1}=n_{2}+n_{3}}}\hspace{-5mm}%
C_{nn_{1}n_{2}n_{3}}^{mm_{1}m_{2}m_{3}}\bar{\alpha}_{n_{1}m_{1}}\alpha
_{n_{2}m_{2}}\alpha _{n_{3}m_{3}}.  \label{eq:Resonant_Eq_resonant}
\end{equation}%
The general expectation is that Eq. (\ref{eq:Resonant_Eq_resonant}) provides
an accurate approximation to the exact evolution equations (\ref%
{eq:Resonant_Eq_v0}), which are tantamount to Eq. (\ref{eq:GP}), over times $%
t\sim 1/g$. Such statements for systems of evolution equations with a finite
number of degrees of freedom can be rigorously proved by elementary methods
\cite{murdock}. At the same time, a mathematically rigorous treatment of
approximating the specific equation (\ref{eq:GP}) by Eq. (\ref%
{eq:Resonant_Eq_resonant}) is given in Ref. \cite{GHT}. Resonant equations
of this type have also been recently employed to study two-component BECs
\cite{Schwinte:2020uqb}, as well as the scattering dynamics governed by NLS
equations when the HO trapping only acts in a single spatial direction \cite%
{Cheng2021}.

As mentioned above, 2D equation (\ref{eq:GP}) has a specific symmetry
structure, which becomes even more manifest in the resonant approximation.
Thus, the evolution defined by Eq. (\ref{eq:Resonant_Eq_resonant}) conserves
the following six quantities \cite{GHT}:
\begin{align}
{N}& =\sum_{nm}\,|{\alpha }_{nm}|^{2},  \label{GP2_N} \\
{M}& =\sum_{nm}m\,|{\alpha }_{nm}|^{2}, \\
{E}& =\sum_{nm}n\,|{\alpha }_{nm}|^{2}, \\
{Z}_{+}& =\sum_{nm}\sqrt{\frac{n+m+2}{2}}\,{\bar{\alpha}}_{n+1,m+1}{\alpha }%
_{nm},  \label{GP2_Z} \\
{Z}_{-}& =\sum_{nm}\sqrt{\frac{n-m+2}{2}}\,{\bar{\alpha}}_{n+1,m-1}{\alpha }%
_{nm}, \\
{W}& =\sum_{nm}\frac{\sqrt{n^{2}-m^{2}}}{2}\,{\bar{\alpha}}_{nm}{\alpha }%
_{n-2,m}.  \label{GP2_W}
\end{align}%
The first two of them are directly inherited from the conservation laws of
Eq. (\ref{eq:GP}) and correspond to the conservation of the wavefunction
norm (number of particles) and angular momentum. The third one is related to
the energy of the linearized version of Eq. (\ref{eq:GP}), which is
conserved by the resonant interactions retained in Eq. (\ref%
{eq:Resonant_Eq_resonant}). Finally, the origin of the three remaining
conserved quantities can be traced back \cite{E} to the \textquotedblleft
breathing modes" of Eq. (\ref{eq:GP}) --- namely those quantities that
evolve periodically for all solutions of the equations of motion. Thus, $%
Z_{+}$ and $Z_{-}$ correspond to the two spatial coordinates of the
center-of-mass of the field configuration described by $\Psi (x,t)$, which
always performs a simple harmonic oscillatory motion \cite{bbbb}, while $W$
corresponds to the Pitaevskii-Rosch breathing mode \cite%
{breathing,qbreathing}.

We finally quote an explicit expression for the interaction coefficients (\ref%
{eq:C_nnnn_mmmm}) through the Laguerre polynomials:
\begin{align}
& C_{n_{1}n_{2}n_{3}n_{4}}^{m_{1}m_{2}m_{3}m_{4}}=\frac{1}{\pi }\left(
\prod_{i=1}^{4}\sqrt{\frac{(\frac{1}{2}(n_{i}-m_{i}))!}{(\frac{1}{2}%
(n_{i}+m_{i}))!}}\right)  \label{CLaguerre} \\
& \times \int_{0}^{\infty }\hspace{-1.5mm}d\rho \,e^{-2\rho }\rho
^{(m_{1}+m_{2}+m_{3}+m_{4})/2}\left( \prod_{i=1}^{4}L_{\frac{n_{i}-m_{i}}{2}%
}^{m_{i}}(\rho )\right) .  \notag
\end{align}%
As the Laguerre polynomials are generated by a simple iterative procedure
that raises their degree (related to the raising operators for the HO),
these interaction coefficients inherit a peculiar discrete structure from
the underlying set of eigenmodes. In particular, they satisfy a set of
simple finite-difference equations \cite{AO,CCEK} linked to the conservation
laws (\ref{GP2_Z}-\ref{GP2_W}). Integrals in Eq. (\ref{CLaguerre}) are known
as \textit{Krein functionals} in mathematical literature \cite{krein}, some
of them featuring prominently in combinatorics \cite{lagcomb1,lagcomb2}. The
discrete nature of the integrals of products of Laguerre polynomials spawns
a profusion of identities for the interaction coefficients that, in turn,
translate into peculiar dynamical patterns in the resonant evolution
described by Eq. (\ref{eq:Resonant_Eq_resonant}), some of which are
considered below.


\section{Restriction of the resonant evolution to subsets of modes}

\label{sec:trunc}

The resonant evolution governed by Eq. (\ref{eq:Resonant_Eq_resonant}) can
be consistently restricted to various subsets of modes in the sense that, if
the initial data excite only modes in a particular subset of this sort, no
modes outside the subset will be excited at later times if the system
evolves according to Eq. (\ref{eq:Resonant_Eq_resonant}). Since the novel
dynamical phenomena that we report in this article unfold within such
restrictions, it is relevant to review this aspect of the model first.

The key aspect of Eq. (\ref{eq:Resonant_Eq_resonant}) that supplies a large
variety of consistent dynamical restrictions is the presence of both the
resonance condition $n+n_{1}=n_{2}+n_{3}$ and the conservation condition for
the angular momentum, $m+m_{1}=m_{2}+m_{3}$ in the summation on the
right-hand side. As a result, if $(n_{1},m_{1})$, $(n_{2},m_{2})$ and $%
(n_{3},m_{3})$ satisfy $an_{j}+bm_{j}=c$, for $j=1,2,3$, with arbitrary $a$,
$b$, $c$, then $(n,m)$ satisfies the same equation. Thus, if the initial
state excites exclusively modes $\alpha _{nm}$ with $an+bm=c$, the time
derivatives of $\alpha _{nm}$ in Eq. (\ref{eq:Resonant_Eq_resonant}) vanish
for all modes $\alpha _{nm}$ with $an+bm\neq c$, hence those modes will
permanently keep zero values. One can thus consistently restrict the
resonant evolution to any straight line traversing the mode tower in Fig.~%
\ref{fig:Diagram_nm}, a few such restrictions having been highlighted in
that figure.

The restrictions reduce the number of indices of modes $\alpha _{nm}$ from
two to one, as $n=n(l)$ and $m=m(l)$ are now linear functions of the new
mode index $l$, hence one can introduce $\alpha _{l}\equiv \alpha
_{n(l)m(l)}$. The two conditions $n+n_{1}=n_{2}+n_{3}$ and $%
m+m_{1}=m_{2}+m_{3}$ in Eq. (\ref{eq:Resonant_Eq_resonant}) now coalesce
into a single resonance condition, $l+l_{1}=l_{2}+l_{3}$, and, under the
adopted mode restriction, one can rewrite Eq. (\ref{eq:Resonant_Eq_resonant}%
) as
\begin{equation}
i\dot{\alpha}_{l}=\hspace{-2mm}\sum_{l+k=i+j}\hspace{-2mm}C_{lkij}\bar{\alpha%
}_{k}\alpha _{i}\alpha _{j}.
\end{equation}%
There is a further dynamical restriction that one can impose, namely, if
the initial state only excites $\alpha _{l}$ with $l=p\mod q$, then no $%
\alpha _{l}$ with $l\neq p\mod q$ will get excited. Thus, not only can one
restrict to straight lines within the mode tower, as in Fig.~\ref%
{fig:Diagram_nm}, but it is also possible to further restrict the evolution
to any regular 1D lattice of modes positioned along any given straight line.

The full variety of such restrictions is conveniently characterized by the
two lowest modes $(n_{0},m_{0})$ and $(n_{1},m_{1})$. Once these modes
have been selected, the full set of modes $\alpha _{l}\equiv \alpha
_{n(l)m(l)}$ that participate in the evolution is defined by $%
n(l)=n_{0}+(n_{1}-n_{0})l$ and $m(l)=m_{0}+(m_{1}-m_{0})l$, and the
evolution equation can be written as
\begin{equation}
i\dot{\alpha}_{l}=\sum_{k=0}^{\infty }\sum_{i=0}^{l+k}C_{lki,l+k-i}\bar{%
\alpha}_{k}\alpha _{i}\alpha _{l+k-i},  \label{eq:Resonant_Eq_resonant_nmkl}
\end{equation}%
where we have converted the resonance condition $l+k=i+j$ into an explicit
specification of the summation ranges. The interaction coefficients in this
restricted resonant system are inherited from Eq. (\ref{CLaguerre}) as $%
C_{lkij}\equiv C_{n(l)n(k)n(i)n(j)}^{m(l)m(k)m(i)m(j)}$. Note that the
interaction coefficients satisfy $C_{lkij}=C_{lkji}=\bar{C}%
_{ijlk}$.

Within the restriction given by Eq. (\ref{eq:Resonant_Eq_resonant_nmkl}),
the conserved quantities (\ref{GP2_N}-\ref{GP2_W}) reduce to a smaller set. 
The first quantity is directly carried over to Eq. (\ref%
{eq:Resonant_Eq_resonant_nmkl}), while the second and third ones merge into
a single conserved quantity, due to the linear relation between $n$ and $m$
imposed by the mode restriction. As a result, one is left with two
quantities conserved by Eq. (\ref{eq:Resonant_Eq_resonant_nmkl}):
\begin{equation}
N=\sum_{l=0}^{\infty }|\alpha _{l}|^{2},\qquad E=\sum_{l=1}^{\infty
}l\,|\alpha _{l}|^{2}.  \label{NEcons}
\end{equation}%
The symmetry transformations corresponding to these conservation laws are
\begin{equation}
\alpha _{l}\rightarrow e^{i\eta }\alpha _{l}\quad \mbox{and}\quad \alpha
_{l}\rightarrow e^{i\theta l}\alpha _{l},  \label{transform}
\end{equation}%
where $\eta $ and $\theta $ are $l$-independent parameters. As to the
remaining three conservation laws (\ref{GP2_Z}-\ref{GP2_W}), they degenerate
to identical zeros for generic straight-line restrictions in Fig.~\ref%
{fig:Diagram_nm}. Exceptions to this rule are given by restrictions along
diagonal lines directed at $45^{\circ }$ in Fig.~\ref{fig:Diagram_nm} (the
Landau levels), that retain either $Z_{+}$ or $Z_{-}$ \cite{BBCE,BBCE2}, or
vertical lines (angular momentum levels) that retain nonzero $W$ \cite{BBCE2}%
. The surviving conservation laws have strong consequences for these
specific restrictions \cite{BBCE,BBCE2}, but they are not present in the
case of generic restrictions that will be our focus below.

In our treatment, we are mainly concerned with restrictions corresponding to
straight lines in Fig.~\ref{fig:Diagram_nm} forming angles $>45^{\circ }$
with the horizontal axis. Restrictions to straight lines at angles $<45^{\circ }
$ are certainly possible, but they are less relevant in relation to topics of
energy transfer, since these restrictions necessarily involve only a finite
set of modes.

Under each specific mode restriction described by Eq. (\ref%
{eq:Resonant_Eq_resonant_nmkl}), we will mostly work with the \emph{two-mode
initial data},
\begin{equation}
|\alpha _{0}(0)|\neq 0,\quad |\alpha _{1}(0)|\neq 0,\quad |\alpha _{l\geq
2}(0)|=0.  \label{eq:two_mode_initial_data}
\end{equation}%
In terms of the original mode tower of Fig.~\ref{fig:Diagram_nm}, this
corresponds to exciting two modes $(n_{0},m_{0})$ and $(n_{1},m_{1})$ and
tracking the subsequent evolution. Such two-mode initial data are the
simplest setup leading to nontrivial evolution, as single-mode initial data
never induce any energy exchange between the modes in the course of the
resonant evolution defined by Eq. (\ref{eq:Resonant_Eq_resonant}). The
dynamical trajectories starting with two-mode initial data can, on the other
hand, display very elaborate behaviors. They have been often studied in the
context of resonant systems of the type of Eq. (\ref%
{eq:Resonant_Eq_resonant_nmkl}), not necessarily related to the present
setup, and have been seen to result, in different situations, in perfectly
periodic evolution of the mode amplitude spectrum \cite%
{BBCE,BBCE2,AO,CF,BEF,GG,CEL}, turbulent phenomena \cite{GG,BE,BMR} and
FPU-like approximate energy returns \cite{FPU,BCE}. Our purpose in the
following two sections is to examine the energy transfer patterns resulting
from two-mode initial data specifically in the context of the resonant
system (\ref{eq:Resonant_Eq_resonant}), derived from the NLS/GP equation
with the isotropic HO potential (\ref{eq:GP}), and identify FPU-like
recurrencies in this setting, as well as special two-mode initial data for
which the energy transfer is completely blocked by the structure of the
interaction coefficients (\ref{CLaguerre}).


\section{Dynamical revivals}

\label{sec:fpu}

To put things in perspective, we pause for a moment and discuss what kind of
behavior one may expect from systems of the form of Eq. (\ref%
{eq:Resonant_Eq_resonant_nmkl}), starting with the two-mode initial
conditions (\ref{eq:two_mode_initial_data}), before specializing to the
actual values of the interaction coefficients $C$ arising from the NLS
equation (\ref{eq:GP}). Initially, the energy leaves the two lowest modes
and spreads to the higher ones, which is often described as a
\textquotedblleft turbulent cascade", as the energy flows towards shorter
wavelengths. The strength of the turbulent cascade varies widely depending
on the precise set of interaction coefficients $C$ \cite{meloturbu}. In
extreme cases, the cascade leads to formation of a power-law spectrum in a
finite time, known as the \textquotedblleft finite-time turbulent blowup"
\cite{BMR,BE}. More conventionally, as is the case for the low-dimensional
NLS equations that we address here, the direct cascade only lasts over a
finite time period, whereupon it halts and turns into a reverse cascade
driving the energy back to the low-lying modes. Note that, due to the
presence of the doublet of conservation laws (\ref{NEcons}), the direct
cascade is itself of a dual nature \cite{dual}, meaning that there is a
simultaneous flow of energy both to higher and lower modes, since it is the
only way to ensure that both quantities (\ref{NEcons}) are conserved.

The efficiency of the reverse cascade also varies from one system to another
and among different initial conditions. For some specific mode restrictions
given by Eq. (\ref{eq:Resonant_Eq_resonant}), such as the Landau-level \cite%
{BBCE,BBCE2} and fixed-angular-momentum \cite{BBCE2} restrictions
highlighted in Fig.~\ref{fig:Diagram_nm}, if the evolution starts with the
two-mode initial data (\ref{eq:two_mode_initial_data}), the direct cascade
is followed by a reverse one that  brings the mode energies back exactly to
the initial configuration. The process then simply repeats itself
periodically. Such perfect dynamical recurrencies are of course highly
non-generic, and they are mandated by the enhanced symmetry structures \cite%
{AO,E,MO} that operate within these specific mode restrictions.

For more generic mode restrictions in Fig.~\ref{fig:Diagram_nm}, which do not
go vertically or diagonally at $45^{\circ }$, there are no reasons to expect
that the reverse cascade will perfectly restore the initial mode energy
distribution, and, indeed, numerical simulations of Eq. (\ref%
{eq:Resonant_Eq_resonant_nmkl}) show that such perfect restoration does not
occur. At the same time, one observes that, for many specific restrictions,
the mode energy distribution at the bottom of the reverse cascade comes very
close to the initial configuration (\ref{eq:two_mode_initial_data}), very
much in the spirit of the FPU phenomena \cite{FPUorg,FPUrev} in nonlinear
oscillator chains. Reporting such dynamical behaviors is one of the main
strands of our presentation. Below we give a heuristic argument to explain
why such FPU-like behaviors may be generically expected in systems of the
form of Eqs. (\ref{eq:Resonant_Eq_resonant_nmkl}), although the actual precision
of the revival by the reverse cascade of the original amplitude spectrum
cannot be accurately determined from such arguments, requiring case-by-case
numerical simulations. Note that the direct-reverse cascade sequences
continue past the first turbulent oscillation that we have outlined above,
and later reverse cascades may restore the initial amplitude spectrum with
an even better precision than the first one, which is also what had happens
in the original FPU setup.

In what follows, we choose two modes $(n_{0},m_{0})$ and $(n_{1},m_{1})$, as
in the previous section, and launch simulations of Eq. (\ref%
{eq:Resonant_Eq_resonant}) with only these two modes excited in the initial
state. As per discussion in the previous section, the subsequent evolution
excites only modes positioned along the straight line passing through the points $%
(n_{0},m_{0})$ and $(n_{1},m_{1})$ in Fig.~\ref{fig:Diagram_nm}, for which
the restricted resonant equation (\ref{eq:Resonant_Eq_resonant_nmkl}) may be
applied, where the two initially excited modes are simply relabelled as mode
0 and mode 1. To quantify the FPU-like phenomena, we measure the
contribution to the \textquotedblleft particle number" $N$ in Eq. (\ref%
{NEcons}) from modes higher than those corresponding to $l=1$:
\begin{equation}
\Delta (\tau )\equiv \sum_{l=2}^{\infty }|\alpha _{l}(\tau )|^{2}.
\label{eq:Delta_Energy}
\end{equation}%
Initially, this is zero by construction, then it starts growing in the
course of the first direct cascade, and we subsequently track the minima of
this function at later times, brought about by the reverse cascades.  The ratio $%
\Delta _{\min }/\Delta _{\max }$ of the minimum and maximum of $\Delta $
provides a characterization of the prominence of the FPU phenomena.
\begin{figure}[t]
\includegraphics[width=%
\columnwidth]{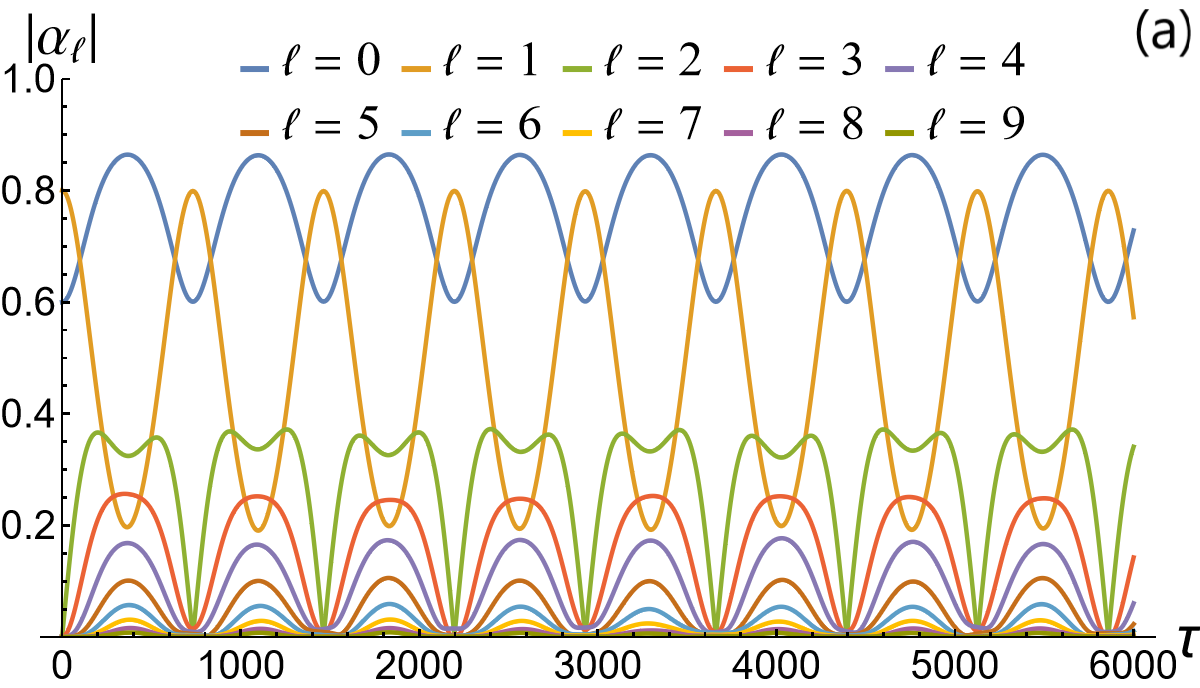}%
\par
\par
\includegraphics[width=%
\columnwidth]{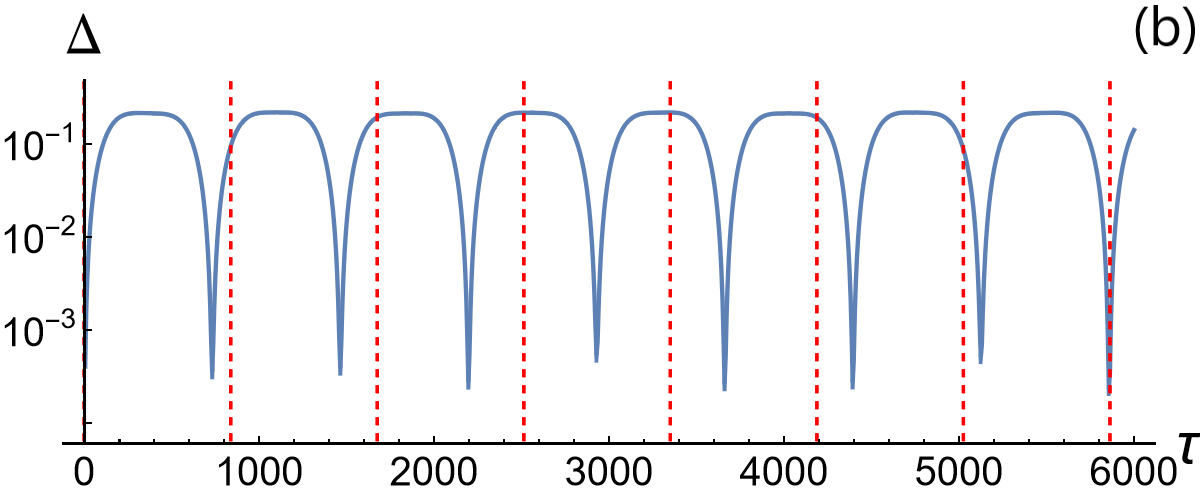}
\par
\par
\includegraphics[width=%
\columnwidth]{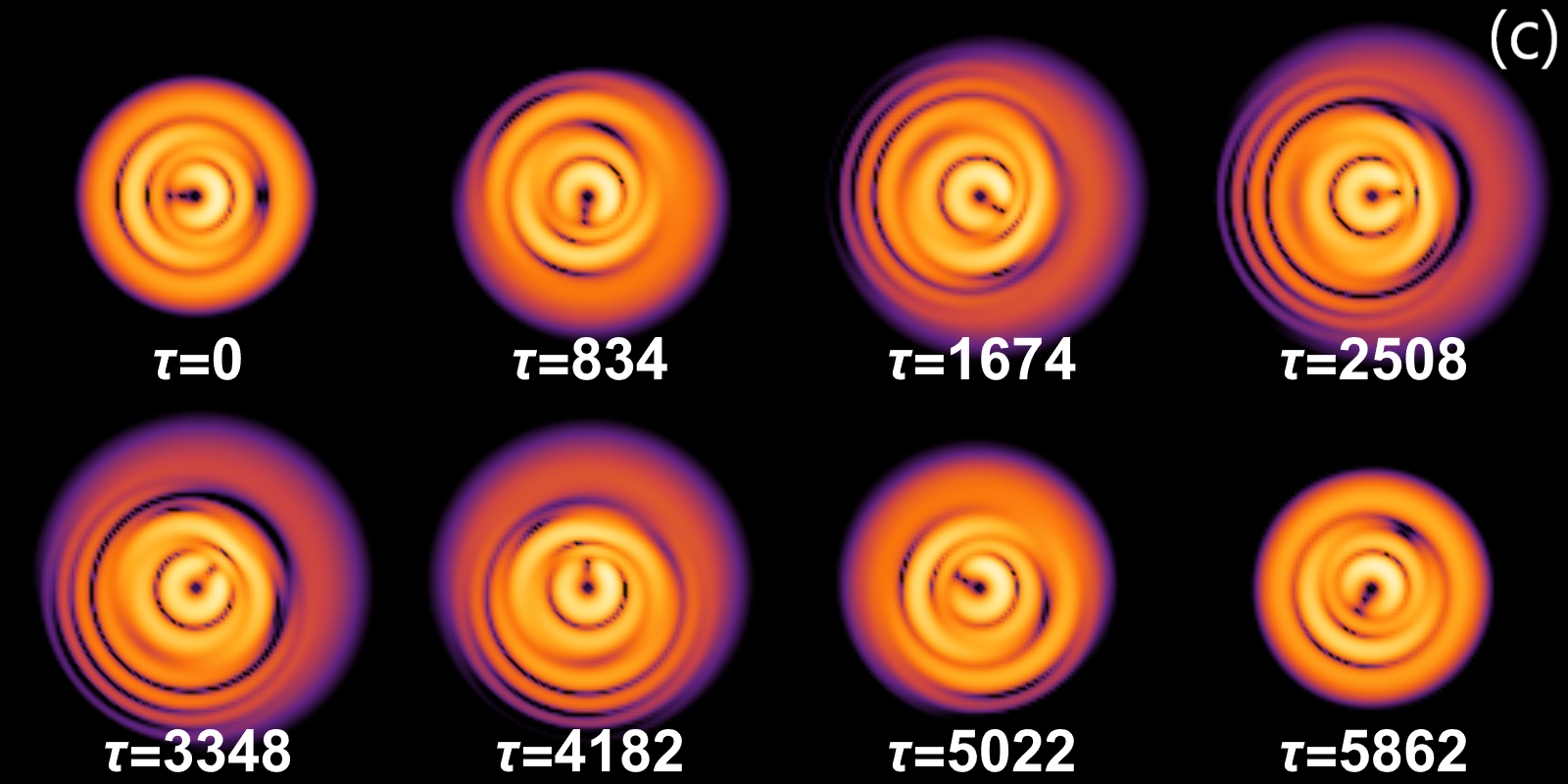}
\par
\par
\includegraphics[width=\columnwidth]{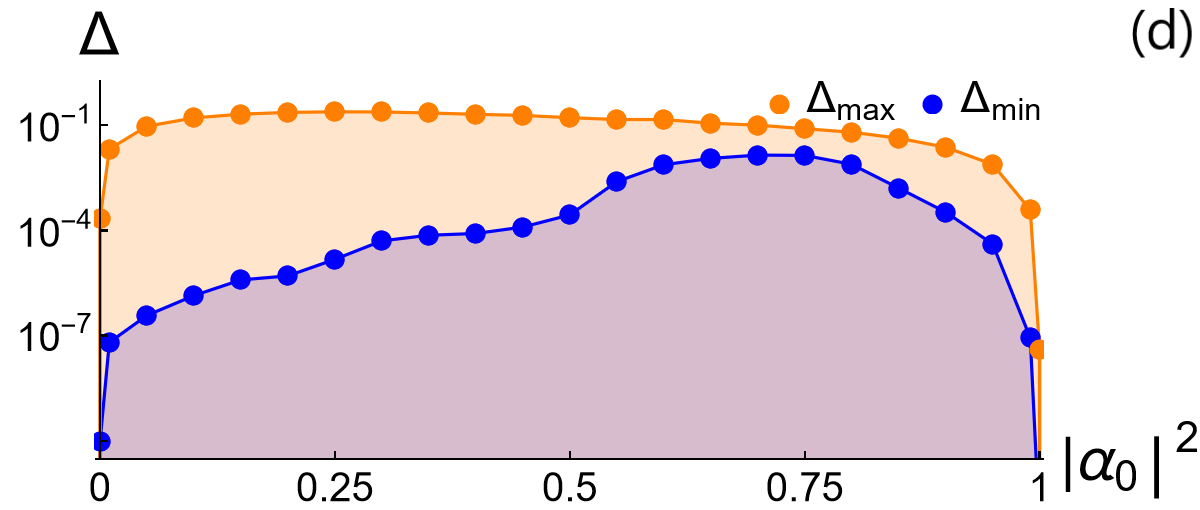}%
\vspace{-5mm}
\par
\vspace{2mm}
\caption{{\protect\small The upper plot: the evolution of $|\protect\alpha %
_{l}|$. The second plot: the contribution to $N$ from modes with }$%
{\protect\small l>1}${\protect\small , defined as per Eq. (\protect\ref%
{eq:Delta_Energy}). The third plot: the position-space representation of $|%
\protect\psi |^{2}$ at times highlighted by red dashed lines in the middle
plot. The initial configuration consists of the two-mode initial data (%
\protect\ref{eq:two_mode_initial_data}). The two modes in the initial state
in these plots are chosen as $(n_{0},m_{0})=(4,-2)$ and $(n_{1},m_{1})=(9,-3)
$ with initial values $\protect\alpha _{0}=0.6$ and $\protect\alpha _{1}=0.8$%
, such that Eq. (\protect\ref{NEcons}) gives $N=1$. The lower plot: the
maximum and minimum values of $\Delta (\protect\tau )$ over a succession of $%
50$ direct-reverse cascades (with the first direct cascade excluded) for the
restriction given in the upper plots but different distributions of the
initial energy between modes 0 and 1 (with $N=1$).}}
\label{fig:Energy_per_Mode}
\end{figure}

We note that the dynamics of two-mode initial data (\ref%
{eq:two_mode_initial_data}) is determined by the initial amplitudes $|\alpha
_{0}|$ and $|\alpha _{1}|$, as the phases of these two modes can be rotated
arbitrarily by the symmetry transformations (\ref{transform}). In the position
space picture, such phase changes are seen as a rotation of the wavefunction density
around the origin. Thus, if a perfect return to the initial configuration of
$|\alpha _{0}|$ and $|\alpha _{1}|$ occurs, one sees a rotated version of
the initial configuration. The subsequent evolution will, of course, repeat
the first-pass revival as transformations (\ref{transform}) are symmetries
of the resonant evolution (\ref{eq:Resonant_Eq_resonant_nmkl}), while the
spatial rotations are a symmetry of the original PDE (\ref{eq:GP}). If
approximate (rather than perfect) revivals of the initial two-mode
configuration occur, this scenario will still be in operation, up to small
distortions.

Figure~\ref{fig:Energy_per_Mode} shows a representative example produced by
the simulations, while a few extra plots are given in Appendix~\ref{app:num}%
. We observe that the evolution of $|\alpha _{l}(\tau )|$ displays
oscillations in the form of direct-reverse cascade sequences, and the
energy, initially located in modes $0$ and $1$, is transferred to higher
modes, followed by accurate returns. In the case shown in Fig.~\ref%
{fig:Energy_per_Mode} this process is so accurate that it is difficult to
distinguish it from exact periodicity in a straightforward graphic
presentation. The initial direct cascade spreads the energy appreciably, so
that one gets $\Delta >0.1$, but at later times the initial configuration is
re-assembled with precision as good as $\Delta \sim 10^{-4}$. In the
corresponding picture of the wavefunction density distribution in the
position-space representation, defined in terms of
\begin{equation}
\psi (\tau ,r,\phi )\equiv \sum_{n,m}\alpha _{nm}(\tau )\,\Psi _{nm}(r,\phi
),  \label{eq:Psi_intpic}
\end{equation}%
one sees that the initial distribution is appreciably deformed and then
recovered with a nearly-perfect precision (the picture gets rotated after
this recovery since, while the amplitudes $|\alpha _{nm}|$ return almost exactly
to their initial values, the phases $\arg (\alpha _{nm})$ undergo a drift). Note
that (\ref{eq:Psi_intpic}) is defined in terms of the slow-time evolution on
time scales $t\sim 1/g$. To convert it to the wavefunction (\ref%
{eq:Psi_sum_linear_modes}) of the original equation (\ref{eq:GP}), one must
apply the evolution operator of the linear quantum HO to $\psi (\tau =gt)$.

It is possible to gain further insight into the FPU phenomena in resonant
system (\ref{eq:Resonant_Eq_resonant_nmkl}) by considering initial
conditions where one of the two modes in the initial state (\ref%
{eq:two_mode_initial_data}) dominates. This analysis is similar to that
developed in Ref. \cite{BCE} for related relativistic systems, where more
detailed and in-depth considerations were reported. We choose for the
presentation here the simpler case when the dominant mode is the one labeled
as mode $0$. The direct cascade is expected to be weak in this case, as
follows a posteriori from the consistency of our analysis presented below,
and is easy to verify numerically. In this situation, it is natural to
assume a strong exponential suppression of the spectrum in the form of
\begin{equation}
\alpha _{n}=\delta ^{n}\,q_{n}(\tau ),  \label{SLL0defeq}
\end{equation}%
with a small free parameter $\delta $. One can then treat the evolution of
this form at leading
order in $\delta $, which results in a simplified system,
\begin{equation}
i\dot{q}_{n}=\bar{q}_{0}(t)\sum_{k=0}^{n}C_{n0k,n-k}q_{k}q_{n-k}.
\label{eq:SLL}
\end{equation}%
Unlike the original resonant system (\ref{eq:Resonant_Eq_resonant_nmkl}),
equations with lower $n$ now decouple from the higher ones, and Eq. (\ref%
{eq:SLL}) can be solved recursively mode-by-mode. We can set $q_{0}(0)=1$,
using the scaling symmetry of Eq. (\ref{eq:Resonant_Eq_resonant_nmkl}) of
the form $\alpha (\tau )\rightarrow \lambda \alpha (\lambda ^{2}\tau )$, and
$q_{1}(0)=1$, adjusting the definition of $\delta $. Then, in the framework
of Eq. (\ref{eq:SLL}), the first two modes simply oscillate as
\begin{equation}
q_{0}(t)=e^{-iC_{0000}\tau },\qquad q_{1}(t)=e^{-iC_{0101}\tau },
\end{equation}%
while higher $q_{n}$ satisfy
\begin{equation}
i\dot{q}_{n}-2C_{0n0n}q_{n}=\bar{q}_{0}%
\sum_{k=1}^{n-1}C_{n0k,n-k}q_{k}q_{n-k},  \label{eq:q_n_mode_0_eq}
\end{equation}%
where the right-hand side only depends on $q_{k}$ with ${k<n}$, which have
already been evaluated at the previous iterative steps, hence they simply
provide a source term for the oscillations of $q_{n}$. As a result, each $%
q_{n}$ is a sum of oscillatory terms proportional to $e^{i\Omega \tau }$,
where, for every such term, $\Omega $ is a linear combination of $C_{0k0k}$
with \textit{integer} coefficients, and only for $k\leq n$. For
completeness, we mention that it is in principle possible for $q_{n}$ to
contain secular terms growing like $\tau e^{i\Omega \tau }$ when the
right-hand side of (\ref{eq:q_n_mode_0_eq}) happens to contain a term that
oscillates with frequency $2C_{0n0n}$. Such behavior occurs in specific
situations where instabilities are present, and has no bearing on the bulk
of our considerations.

From the perspective of Eq. (\ref{eq:q_n_mode_0_eq}), FPU phenomena enter
the stage naturally in the following manner. Solving Eq. (\ref%
{eq:q_n_mode_0_eq}) for $q_{2}$, with initial condition $q_{2}(0)=0$, yields
\begin{equation}
|q_{2}|\sim \sin \left( \frac{1}{2}(C_{0000}-4C_{1010}+2C_{2020})\tau
\right) ,
\end{equation}%
which shows that $q_{2}$ periodically vanishes. When this happens, the
energy is entirely concentrated in modes $0$ and $1$, except for the
contributions in modes $3$ and higher, suppressed by $\delta ^{6}$. This is
simply a reflection of a single direct-reverse cascade return in the limit
of small $\delta $. More accurate returns may occur at later times. The
point is that, due to the special nature of the interaction coefficients (%
\ref{CLaguerre}) expressed through a highly structured family of orthogonal
polynomials, many of these coefficients are rational numbers. We exhibit
below specific mode restrictions leading from Eq. (\ref%
{eq:Resonant_Eq_resonant}) to Eq. (\ref{eq:Resonant_Eq_resonant_nmkl}) such
that \textit{all} coefficients $C_{0n0n}$ are rational numbers in units of $%
C_{0000}$. If all $C_{0n0n}$ are rational in units of $C_{0000}$, then the
solutions for all $q_{n}$ are superpositions of oscillations with rational
frequencies, which means that any finite subset of $q_{n}$ has a common
period. For example, it is guaranteed that a moment of time exists when $%
q_{2}$ and $q_{3}$ will return to the initial configuration, where they both
vanish. At that moment, the energy returns to modes $0$ and $1$ with
precision $\delta ^{8}$. Even more broadly, for those mode restrictions
where $C_{0n0n}$ are not all rational, approximate common periods may exist,
securing dynamical returns with enhanced precision, as discussed in Ref.
\cite{BCE}.

We now present a specific simple family of mode restrictions where the
picture outlined above plays a role. In this family, mode $0$ is chosen to
be $(n_{0},m_{0})=(0,0)$ and mode $1$ is any other mode $(n,m)$. Then, mode
number $l$ is $(nl,ml)$. We claim that, under this restriction, the following
relation holds:
\begin{equation}
\frac{C_{0l0l}}{C_{0000}}=\frac{(nl)!}{2^{nl}\left( \frac{1}{2}(n+m)l\right)
!\left( \frac{1}{2}(n-m)l\right) !},  \label{C0l0l}
\end{equation}%
hence all $C_{0l0l}$ have rational values in units of $C_{0000}$. Thus,
synchronization takes place between periods of different modes in Eq. (\ref%
{eq:q_n_mode_0_eq}), providing for returns with improved precision. A
derivation of Eq. (\ref{C0l0l}) is given in appendix~\ref{app:C0l0l}.

Of course, our analysis of the dynamics close to mode $0$ is only valid at
its face value when mode $1$ is strongly suppressed in the initial
configuration. However, it does provide correct intuition in relation to the
direct-reverse cascade sequences produced by Eq. (\ref%
{eq:Resonant_Eq_resonant_nmkl}), and allows one to predict which numbers of
direct-reverse cascades result in particularly accurate returns. These
predictions, furthermore, continue to hold even for initial conditions where
modes $0$ and $1$ carry comparable energies, as has been demonstrated by detailed
analysis of a related resonant system in \cite{BCE}.

Similar considerations are possible for dynamical trajectories dominated by
mode $1$ \cite{BCE}. In that case, instead of relation (\ref{SLL0defeq}),
one assumes a $\delta $-dependence in the form of $\alpha _{0}=\delta q_{0}$%
, $\alpha _{l\geq 1}=\delta ^{l-1}q_{l}$. This specification of $\de$-dependencies 
is consistent in
the sense that, upon the substitution into resonant system (\ref%
{eq:Resonant_Eq_resonant_nmkl}) and taking the limit $\delta \rightarrow
0$, one obtains a simplified and yet nontrivial dynamical system for $q_{l}$%
, which can be analyzed by methods similar to the ones employed above for
solutions dominated by mode $0$. This results in a picture of direct-reverse
cascade sequences, as well as dynamical returns of enhanced precision.
Further details can be recovered from Ref. \cite{BCE}.

The fact that the FPU behaviors (which can be naturally explained in a
quantitative manner for initial conditions dominated by one of the two
modes) persist for initial conditions with comparable mode energies remains
an empirical observation. We have observed such returns for a number of
choices of the initial configurations with comparable energies of the initially
excited modes, as documented by Fig.~\ref{fig:Energy_per_Mode} and further
simulations presented in appendix~\ref{app:num}. It can be seen as a
manifestation of the special character of the 2D NLS/GP equation (\ref{eq:GP}%
) that the FPU behaviors, while being rather generic for initial data
dominated by a single mode, extend, in the case of this equation, to a
broader range of initial conditions. Pronounced FPU phenomena with $\Delta
_{\min }/\Delta _{\max }\sim 10^{-4}$ have been observed in our simulations
for initial data with the following pairs of modes $(n_{0},m_{0})$-$%
(n_{1},m_{1})$: (0,0)-(3,1); (0,0)-(4,2); (0,0)-(5,3); (1,-1)-(4,-2);
(1,-1)-(4,2); (1,-1)-(6,0); (2,0)-(6,0); (2,0)-(5,1); (2,-2)-(5,-1);
(2,-2)-(6,-4); (4,-2)-(9,-3); (4,-2)-(7,-1). These observations suggest that
dynamical recurrencies are not uncommon in the evolution governed by Eq. (%
\ref{eq:Resonant_Eq_resonant}).


\section{Two-mode persistent breathers}

\label{sec:breath}

In our searches for FPU behaviors within various mode restrictions imposed
on Eq. (\ref{eq:Resonant_Eq_resonant}), we have discovered some initial
conditions that lead to an even more dramatic failure of the energy
redistribution among the normal modes. Rather than repeatedly returning to
the vicinity of the original configuration, the energy for such initial
conditions does not get transferred at all, at any time. We have already
mentioned that this property holds in solutions of Eq. (\ref%
{eq:Resonant_Eq_resonant}) for all single-mode initial conditions, but it
does not generically hold for two-mode initial conditions. We have, however,
discovered some special choices of two-mode initial conditions for which no
energy transfer occurs in the course of the resonant evolution of Eq. (\ref%
{eq:Resonant_Eq_resonant}).

The evolution with two-mode initial conditions always unfolds within one of
the mode restrictions defined as per Eq. (\ref{eq:Resonant_Eq_resonant_nmkl}), namely, the
restriction to the modes that fall onto the straight line passing through
the two initially excited modes in Fig.~\ref{fig:Diagram_nm}. The energy
transfer in (\ref{eq:Resonant_Eq_resonant_nmkl}) is blocked whenever the
initial conditions are of the form (\ref{eq:two_mode_initial_data}) and the
interaction coefficient $C_{2011}=0$. Indeed, for the initial conditions of
this form, $\dot{\alpha}_{l}(0)$ with $l\geq 3$ are identically zero since
each term on the right-hand side of (\ref{eq:Resonant_Eq_resonant_nmkl})
involves at least one of the initially vanishing modes $\alpha _{l}$ with $%
l\geq 2$. On the other hand,
\begin{equation}
i\dot{\alpha}_{2}(0)=C_{2011}\,\bar{\alpha}_{0}(0)\,\alpha _{1}^{2}(0).
\end{equation}%
Thus, the inception of energy transfer entirely hinges on $C_{2011}$. If
this coefficient vanishes, then neither mode $2$ nor any higher ones get
excited at the start of the evolution, hence they cannot be excited at any
later stage either, with the evolution perpetually locked within the ansatz (%
\ref{eq:two_mode_initial_data}).

We have identified the following large family of two-mode initial data for
which the corresponding coefficient $C_{2011}$ vanishes:
\begin{equation}
(n_{0},m_{0})=(2i+1,-(2j+1))\quad (n_{1},m_{1})=(2k,0)
\label{eq:stationary_solutions_family_1}
\end{equation}%
with $j\leq i<k$ (to ensure that $(n_{0},m_{0})$ is indeed the lowest mode
of the given restriction, one must supplement this condition with an extra
requirement that $2(i-j)<k$). We provide a (relatively involved) proof for
the vanishing of $C_{2011}$ corresponding to this family in appendix~\ref%
{app:vanish}. We have additionally observed the following more sparse
families involving high-frequency modes that share the same property:
\begin{equation}
(n_{0},m_{0})=(2,0),\quad (n_{1},m_{1})=(k^{2},k)
\label{eq:stationary_solutions_family_2}
\end{equation}%
and
\begin{equation}
(n_{0},m_{0})=(k^{2}-k,-k^{2}+k+2),\quad (n_{1},m_{1})=(k^{2},k)
\label{eq:stationary_solutions_family_3}
\end{equation}%
with $k=2,3,4,...$. Evidently, the sign of the angular momentum can be
flipped in all of these families due to the reflection symmetry.
\begin{figure}[t]
\includegraphics[width=%
\columnwidth]{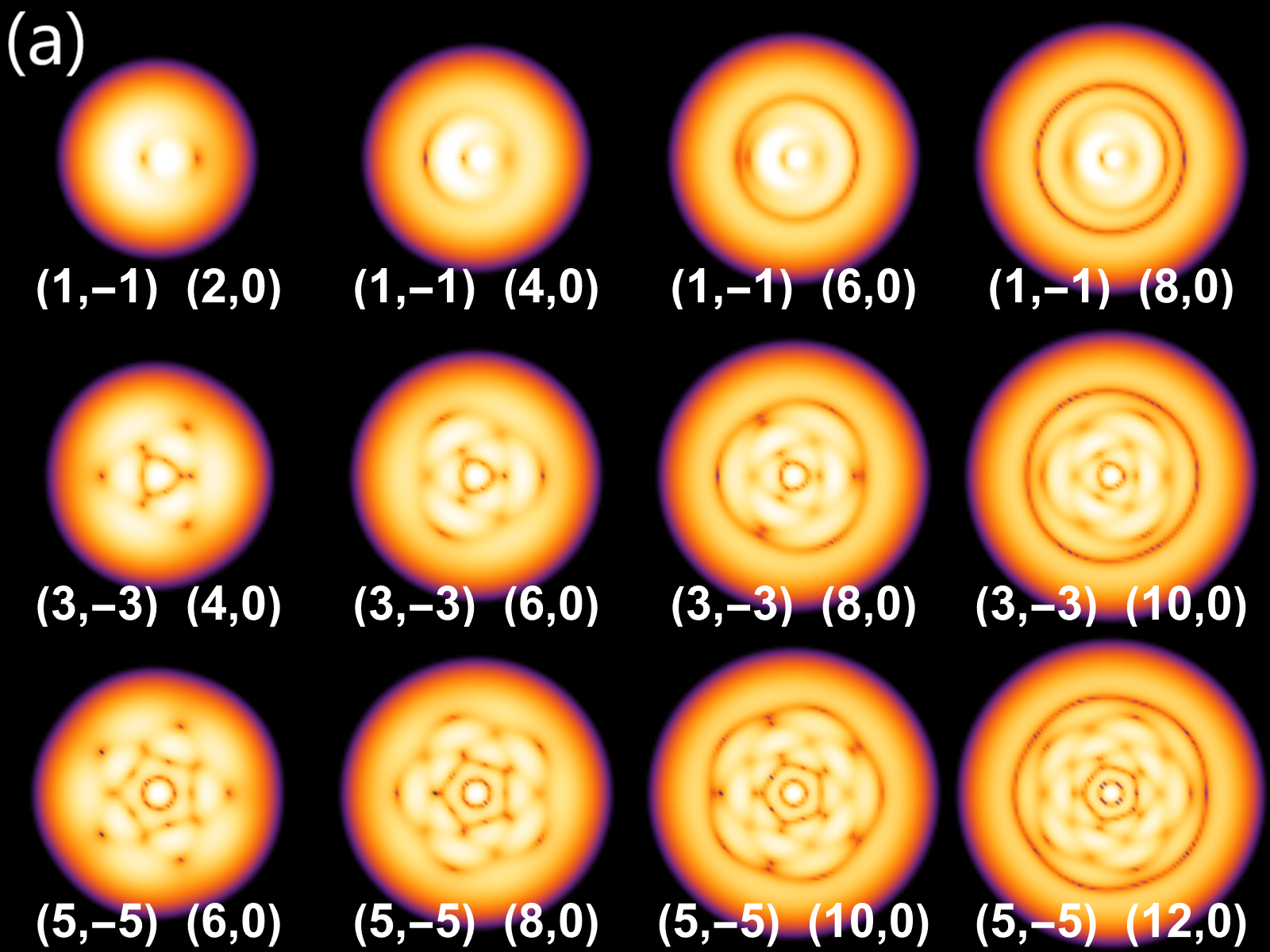} %
\includegraphics[width=%
\columnwidth]{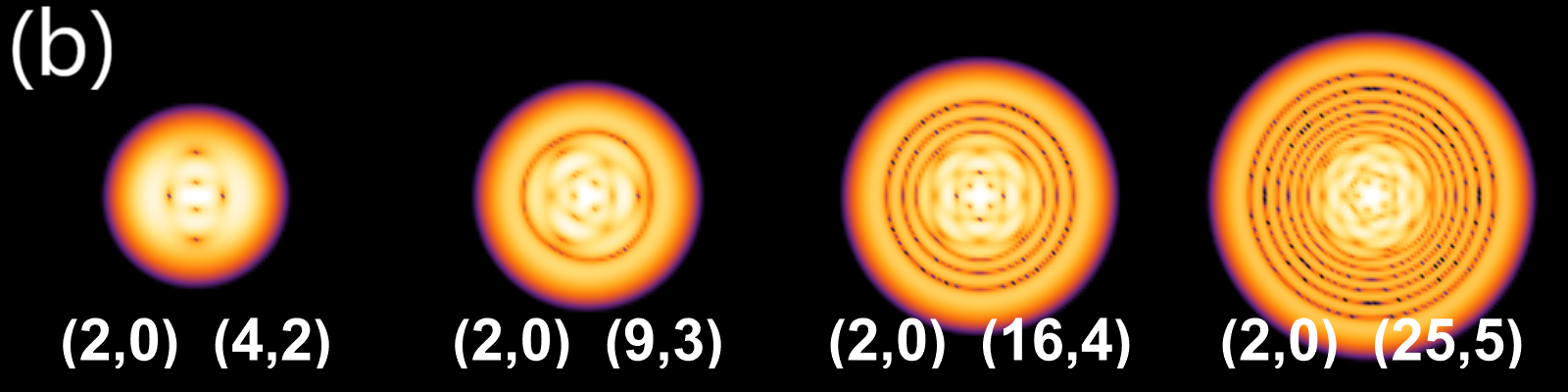}
\caption{{\protect\small The spatial configuration of $|\protect\psi |^{2}$
in the stationary solutions given by Eqs. (\protect\ref%
{eq:stationary_solutions_family_1}) (rows from 1 to 3, (a)) and (\protect\ref%
{eq:stationary_solutions_family_2}) (the lower row, (b)). Each plot specifies two
excited modes $(n_{0},m_{0})$ and $(n_{1},m_{1})$, as indicated by white
numbers.}}
\label{fig:Profiles_family}
\end{figure}
A collection of position space distributions of the wavefunction density
obtained from Eq. (\ref{eq:Psi_intpic}) for these two-mode configurations is
plotted in Fig.~\ref{fig:Profiles_family}. Note that the complex amplitudes
of the two modes activated in the initial state may be prescribed
arbitrarily for each specification of the integer parameters that define the
above two-mode families. While within the resonant approximation (\ref%
{eq:Resonant_Eq_resonant}) these configurations are exactly stationary in
the sense that $|\alpha _{n_{0}m_{0}}|^{2}$ and $|\alpha _{n_{1}m_{1}}|^{2}$
are frozen in time, while all other modes remain zero, the corresponding
initial data in the original PDE (\ref{eq:GP}) will lead to some evolution,
but only in an extremely slow form at small $g$. On time scales $t\gg 1/g$,
the resonant approximation ceases to be accurate and solutions of (\ref%
{eq:GP}) may drift away from these predictions. Yet on time scales $t\sim 1/g
$, the initial data in the form of Eq. (\ref%
{eq:stationary_solutions_family_1}-\ref{eq:stationary_solutions_family_2})
continue tracking the linearized evolution very closely, forming a
long-lived breather-like stationary pattern. This behavior is strongly
non-generic for weakly nonlinear NLS/GP equations (\ref{eq:GP}) with
abundant normal-mode resonances.

To put our finding of two-mode stationary solutions in perspective, we
further remark that resonant systems of the form (\ref%
{eq:Resonant_Eq_resonant_nmkl}) generically possess stationary solutions
without energy transfer. These solutions have spectra given by $\alpha
_{n}=e^{i(\omega +n\lambda )\tau }A_{n}$, where $A_{n}$ are $\tau $%
-independent and depend on $C_{nmkl}$ (upon substituting this ansatz into (%
\ref{eq:Resonant_Eq_resonant_nmkl}), all the $\tau $-dependent factors drop
out, leaving a system of algebraic equations for $\omega $, $\lambda $ and $%
A_{n}$.) Such solutions, however, generically have infinitely many modes
turned on, with a very specific arrangement of amplitudes and phases, and
would be difficult to recreate. The two-mode stationary solutions we report
here are, by contrast, specific to the model we consider, and their
existence relies on the vanishing of particular four-mode couplings, which
is a special feature of the NLS/GP equation (\ref{eq:GP}).


\section{Conclusion}

\label{sec:conclusion}

We have considered the long-term weakly nonlinear evolution of the 2D NLS/GP
equation with the isotropic HO (harmonic oscillator) potential. The analysis
has revealed an array of FPU phenomena and stationary configurations in this
regime. Our findings further highlight the special nature of the
low-dimensional NLS/GP equations with the HO potentials, which previously
exhibited other unusual dynamical behaviors \cite%
{quasiint1,quasiint2,tri,tri1,tri2,BBCE,BBCE2}. There are further physical
connections between such behaviors and obstructions to ergodicity in
low-dimensional many-boson systems in HO potentials \cite%
{atomtherm,atomtherm1,atomtherm2}, as all such phenomena originate from
failures of effective redistribution of energy among the dynamical degrees
of freedom.

It would be interesting to observe the dynamical patterns that we have
displayed here in the evolution of trapped ultracold atomic gases, where HO
potentials are routinely used, while the strength of the interactions can be
adjusted by means of the Feshbach resonances. Detailed control over the
preparation of the initial configurations remains a challenge, but it has
been a focal point in recent research and substantial progress can be seen
in experimental work \cite{tri}.

In the context of nonlinear optics, both our 2D setup (corresponding to the
transverse geometry of the waveguide) and the weakly nonlinear regime are
very natural. In particular, considerations of the coexistence of a large
number of transverse modes when nonlinearities are weak are highly relevant
for the implementation of spatial-division-multiplexing schemes in optical
data-transmission links \cite{spatial}. HO potentials are less common in
this setting, but they may be relevant too \cite{Agrawal}. Note that the
stationary solutions found in section~\ref{sec:breath} are distinguished by
their quasi-linear evolution on long time scales, despite the fact that even
weak nonlinearities could in principle generate large effects through the
resonant interactions.

In a more speculative mode, one can try to employ the FPU phenomena in a
sort of cryptographic scenario, with the communication line passing through
an area where the transmission may be tapped. In this case, one could adjust
the operation mode in such a way that the nonlinearities scramble the signal
in the intermediate region (where the energy distribution of the initial
signal is driven to the excited transverse modes by nonlinear interactions),
and then the original transmission gets reconstructed by an FPU return at
the read-out point. Even within our simple two-mode specification for the
initial states used in section~\ref{sec:fpu}, the relative energy of the two
modes can be used to transmit data, while the overall power of the signal
can be used to ensure that the FPU recurrence takes place precisely at the
specified read-out point.

\section*{Acknowledgments}

A.B. thanks A.~Paredes for useful discussions. A.B. has been supported by the
Polish National Science Centre grant No. 2017/26/A/ST2/00530, O.E. has been
supported by CUniverse research promotion project (CUAASC) at Chulalongkorn
University and by Research Foundation-Flanders through project G006918N. B.A.M. has been
supported by the Israel Science Foundation, through grant No. 1286/17.


\appendix

\section{Additional numerical simulations}

\label{app:num}

We complement the presentation in section~\ref{sec:fpu} with Fig.~\ref%
{fig:Many_plots} containing a compilation of extra plots produced by our
simulations. These plots contain a number of cases where the general
scenario outlined in section~\ref{sec:fpu} is reproduced in detail, as well
as some examples where the FPU returns are less pronounced. There is also a
particular truncation where the single-mode solution supported by mode 1 is
unstable, leading to a breakdown of the picture of strong exponential
suppression in the spectra originating from initial data dominated by one
mode. With all of these exceptions, FPU returns remain rather generic in the
setup of this paper, and we have observed accurate returns in truncations
defined by $(n_{0},m_{0})$-$(n_{1},m_{1})$: (0,0)-(3,1); (0,0)-(4,2);
(0,0)-(5,3); (1,-1)-(4,-2); (1,-1)-(4,2); (1,-1)-(6,0); (2,0)-(6,0);
(2,0)-(5,1); (2,-2)-(5,-1); (2,-2)-(6,-4); (4,-2)-(9,-3); (4,-2)-(7,-1), as
already mentioned in the main text.

\onecolumngrid

\begin{figure}[h]
\includegraphics[width=5cm]{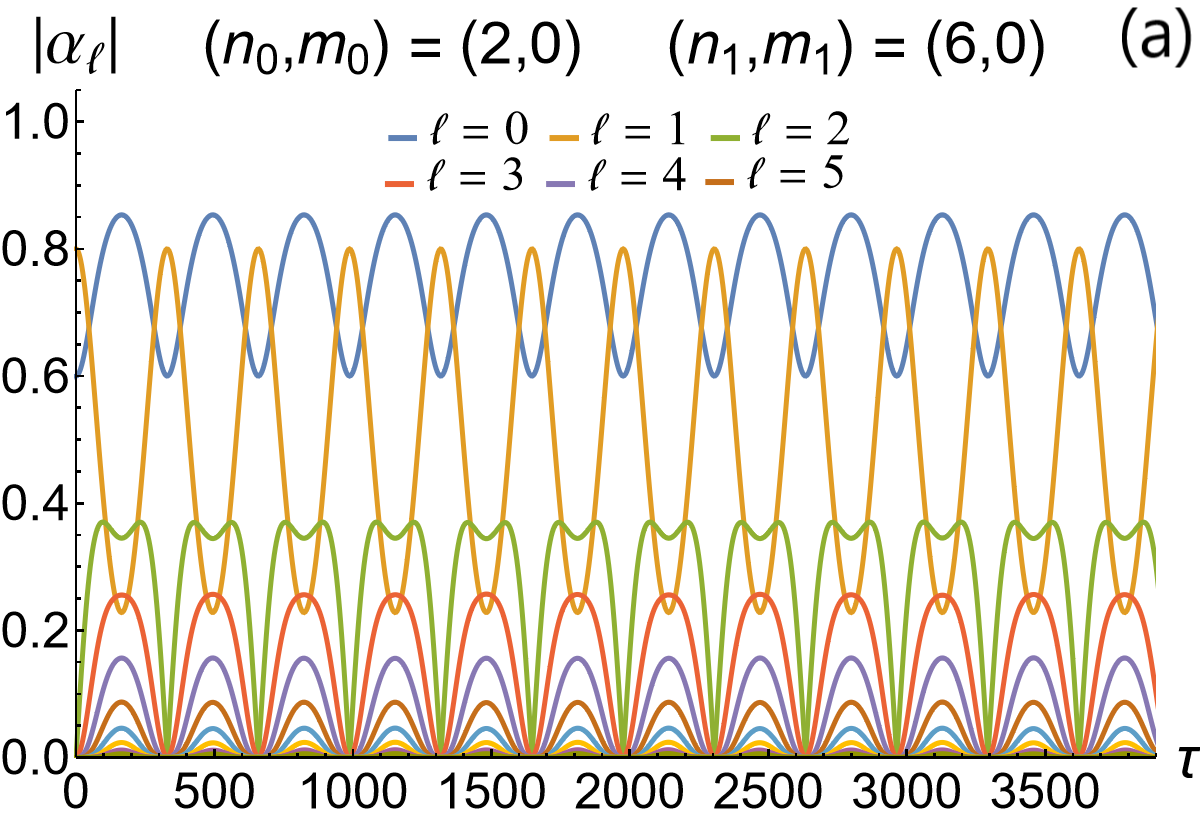}\hspace{0.5cm} %
\includegraphics[width=5cm]{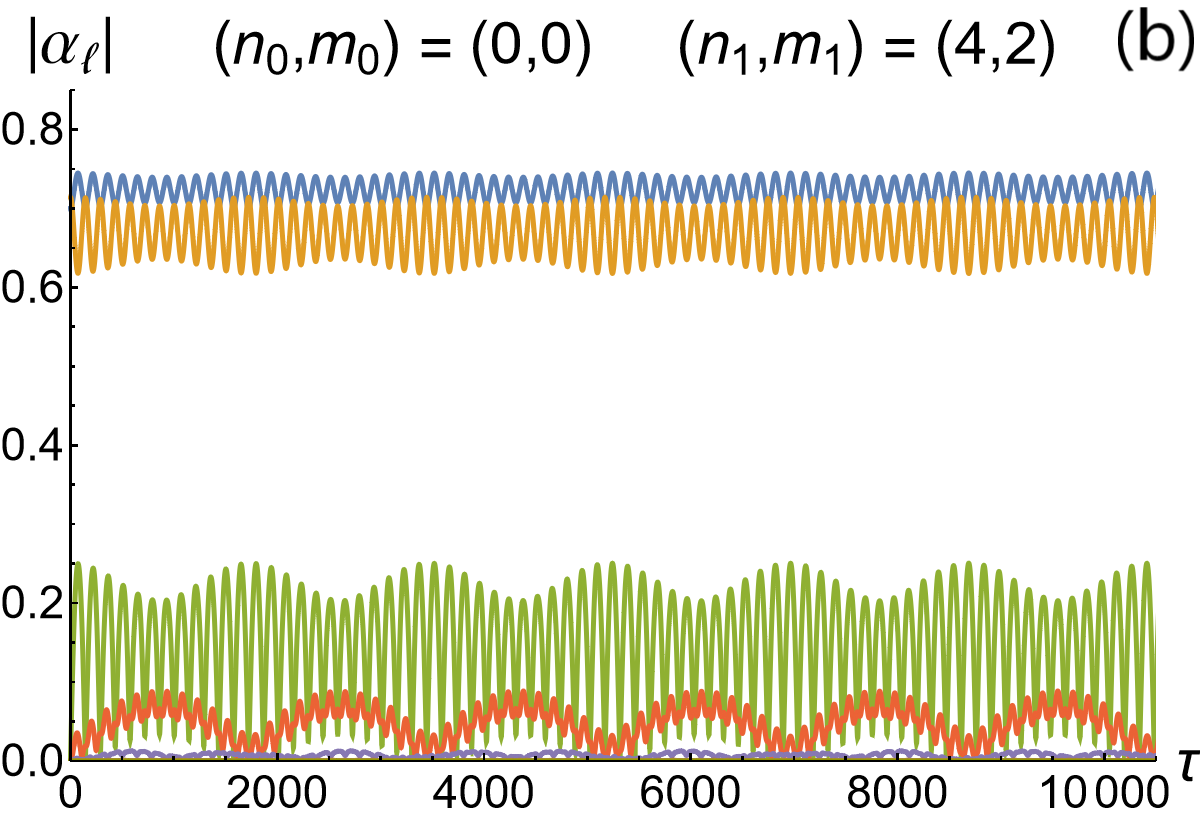}\hspace{0.5cm} %
\includegraphics[width=5cm]{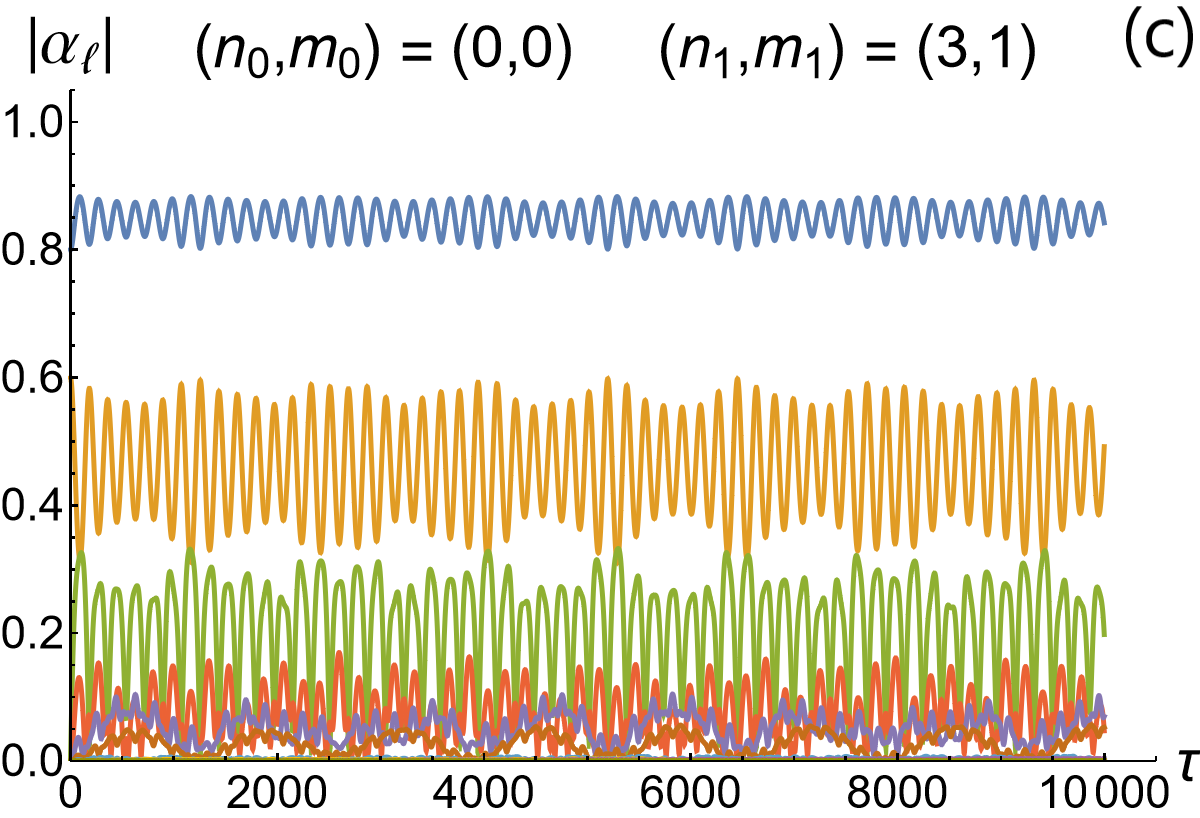}
\par
\vspace{0.5cm}
\par
\includegraphics[width=5cm]{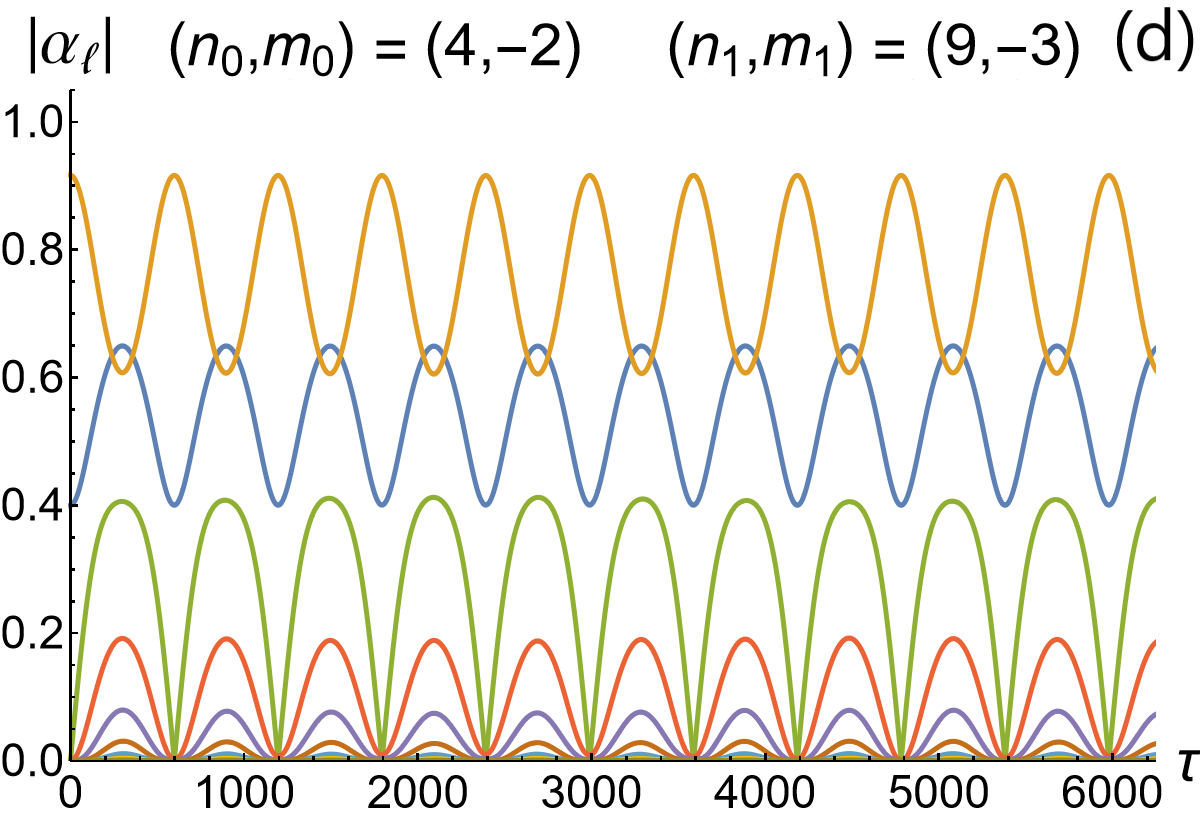}\hspace{0.5cm} %
\includegraphics[width=5cm]{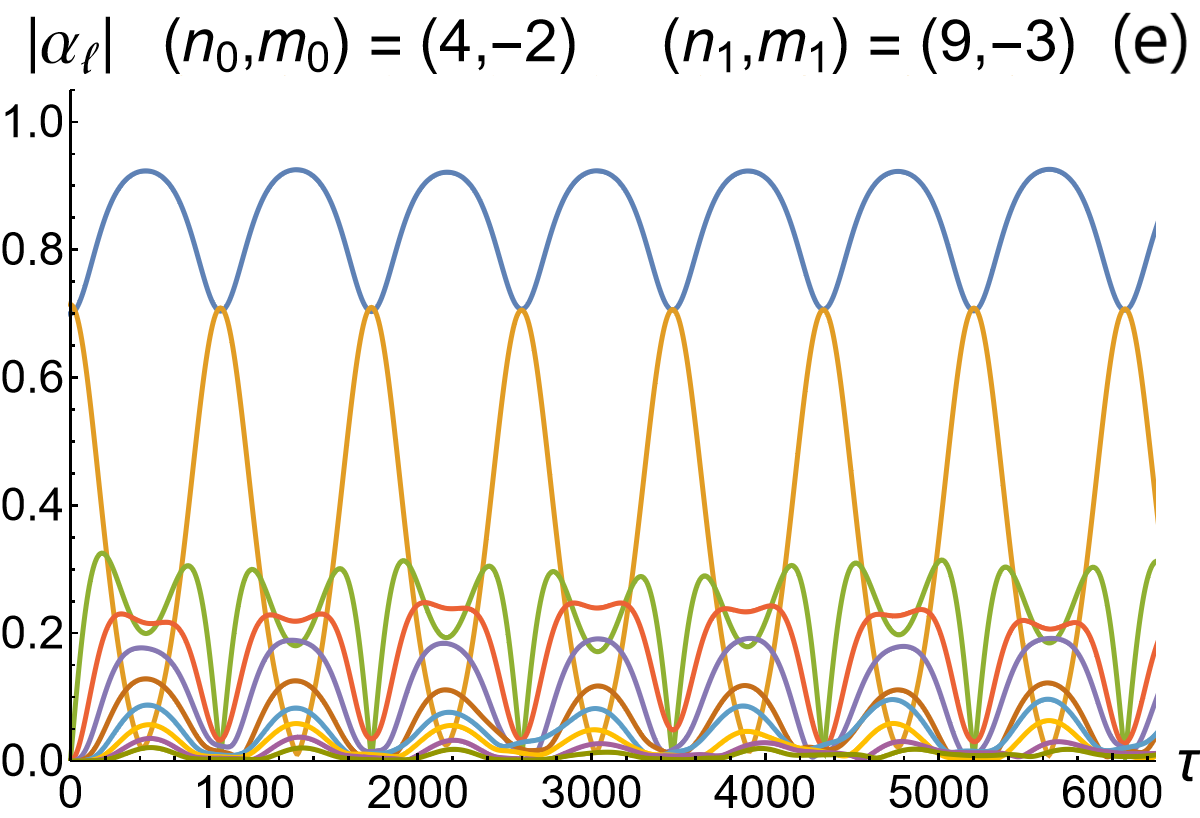}\hspace{0.5cm} %
\includegraphics[width=5cm]{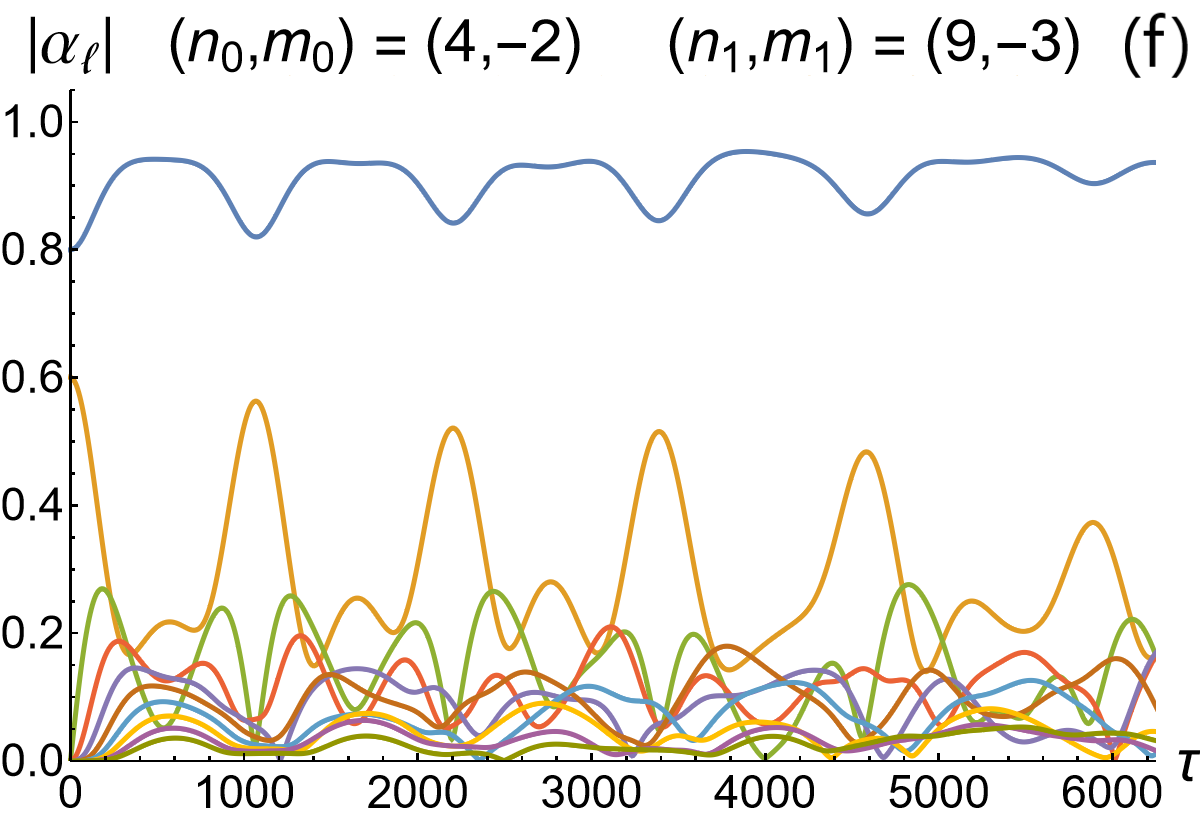}
\par
\vspace{0.5cm}
\par
\includegraphics[width=5cm]{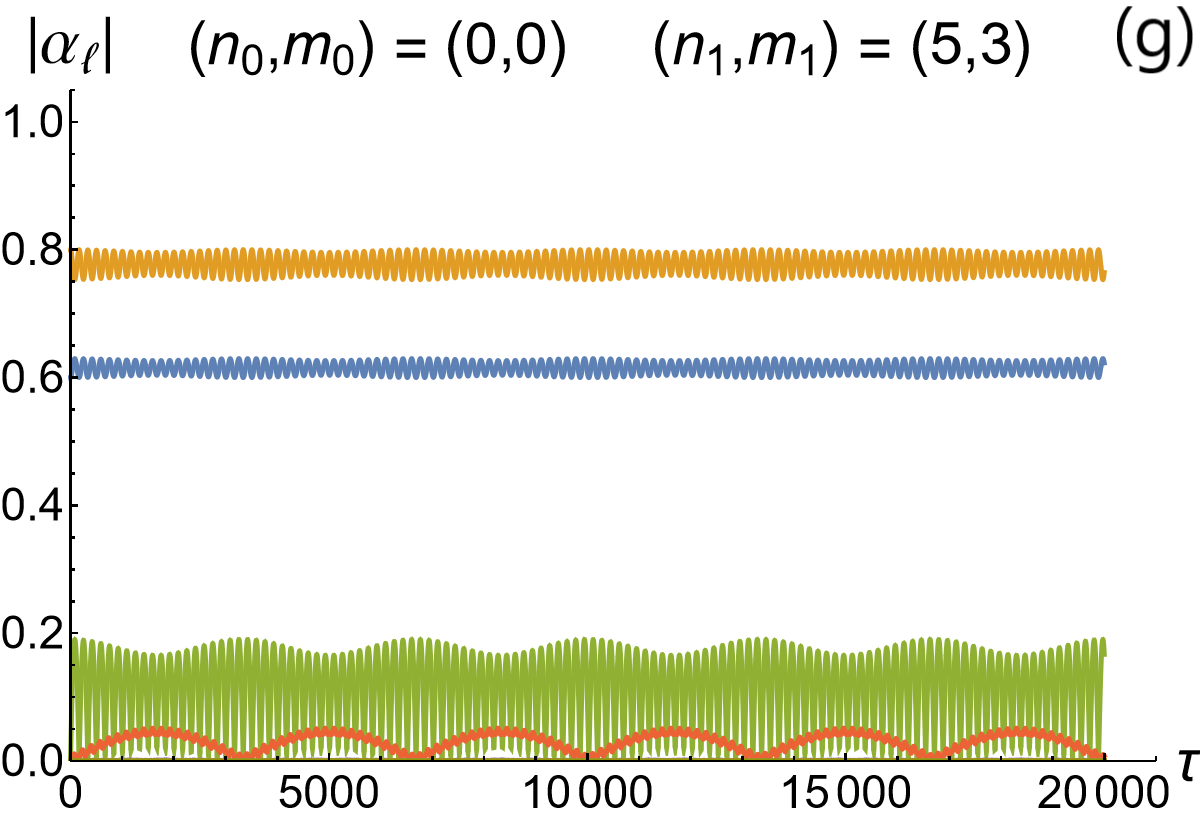}\hspace{0.5cm} %
\includegraphics[width=5cm]{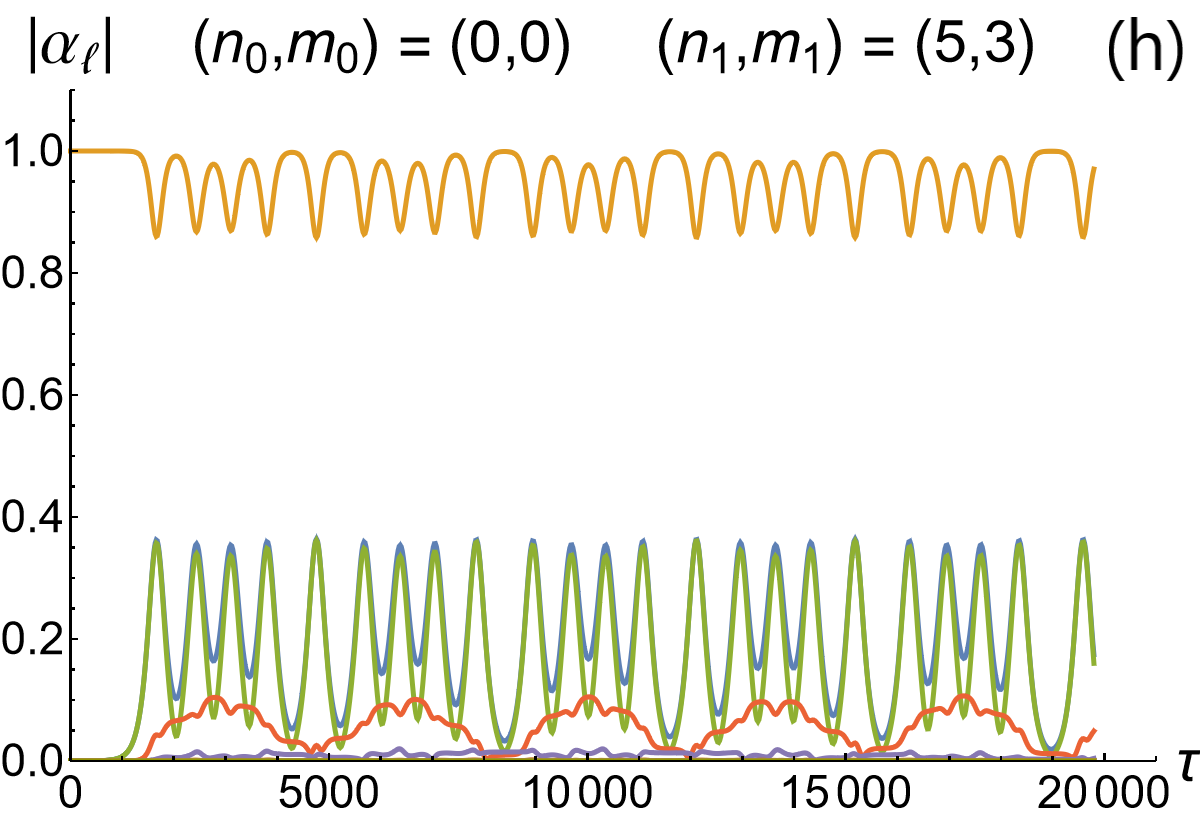}\hspace{0.5cm} %
\includegraphics[width=5cm]{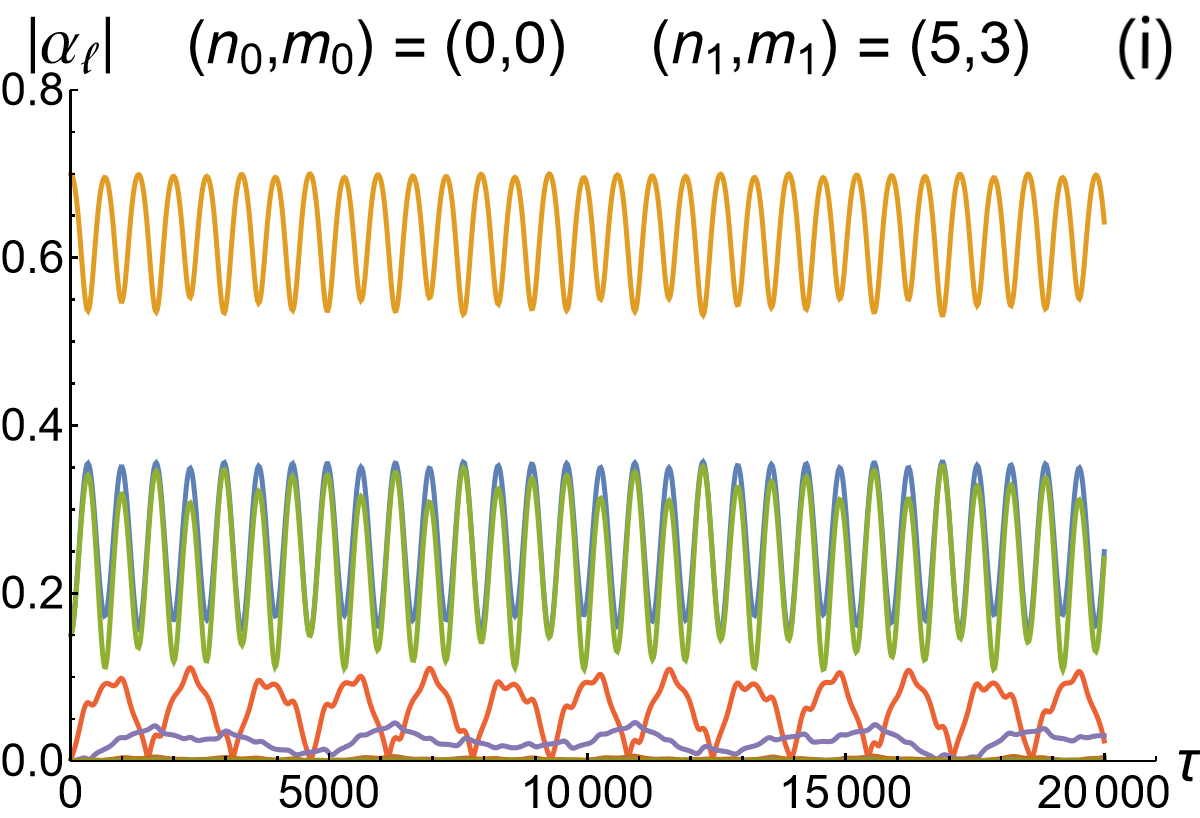}
\par
\vspace{0.5cm}
\par
\vspace{-2mm}
\caption{{\protect\small \emph{Upper row:} The evolution within different
mode restrictions (specified in each plot) that exhibit accurate energy
returns for some two-mode initial data; from left to right $(\protect\alpha %
_{0},\protect\alpha _{1})=(0.6,0.8)$, $(\protect\alpha _{0},\protect\alpha %
_{1})=(0.7,0.7)$, and $(\protect\alpha _{0},\protect\alpha _{1})=(0.8,0.6)$.
\emph{Middle row:} The evolution under a specific mode restriction for
different initial distributions of energy, from left to right $(\protect%
\alpha _{0},\protect\alpha _{1})=(0.4,0.92)$, $(\protect\alpha _{0},\protect%
\alpha _{1})=(0.7,0.7)$, and $(\protect\alpha _{0},\protect\alpha %
_{1})=(0.8,0.6)$. Even when some initial configurations of the two lowest
modes display accurate energy returns, for others such behavior may be much
less pronounced. \emph{Lower row:} This mode restriction demonstrates a
wider range of behaviors than what we have discussed in the main
presentation. In the first plot the two-mode initial data $(\protect\alpha %
_{0},\protect\alpha _{1})=(0.8,0.6)$ results in accurate FPU returns. The
second plot shows that the single mode 1 is unstable because small
perturbations like $(\protect\alpha _{0},\protect\alpha _{1})=(\protect%
\delta ,1)$ with $\protect\delta \ll 1$ drive the system away from this
configuration; however, even if the energy does not accurately return close
to the initial distribution we observe that there are accurate returns
around a new reference configuration. Finally, the third plot shows that
three-mode initial data, specifically $(\protect\alpha _{0},\protect\alpha %
_{1},\protect\alpha _{2})=(0.15,0.7,0.15)$, can exhibit accurate FPU
returns. Mode labelling in the plots is identical to that for the first plot
shown.}}
\label{fig:Many_plots}
\end{figure}


\section{Proof of Eq. (\protect\ref{C0l0l})}

\label{app:C0l0l}

To prove our claim in Eq. (\ref{C0l0l}) that $C_{0l0l}$ are rational in
units of $C_{0000}$, specifically
\begin{equation}
\frac{C_{0l0l}}{C_{0000}}=\frac{(nl)!}{2^{nl}\left( \frac{1}{2}(n+m)l\right)
!\left( \frac{1}{2}(n-m)l\right) !},  \label{C0l0l_v2}
\end{equation}%
one writes, from Eq, (\ref{CLaguerre}),
\begin{equation}
C_{0l0l}=\frac{1}{\pi }\frac{\left( \frac{1}{2}(n-m)l\right) !}{\left( \frac{%
1}{2}(n+m)l\right) !}\int_{0}^{\infty }\hspace{-1.5mm}d\rho \,e^{-2\rho
}\rho ^{ml}\left[ L_{\frac{(n-m)l}{2}}^{ml}(\rho )\right] ^{2}.
\end{equation}%
The integrand is decomposed in a sum of Laguerre polynomials using \cite%
{BookIntegrals}
\begin{equation}
\left( L_{n}^{m}(\rho )\right) ^{2}=\frac{(n+m)!}{2^{2n}n!}\sum_{j=0}^{n}%
\binom{2n-2j}{n-j}\frac{(2j)!}{j!(j+m)!}L_{2j}^{2m}(2\rho ),
\end{equation}%
and after that, individual integrals are evaluated using \cite{BookIntegrals}
\begin{equation}
\int_{0}^{\infty }x^{\gamma -1}L_{n}^{m}(x)e^{-x}dx=\frac{(\gamma
-1)!(n+m-\gamma )!}{n!(m-\gamma )!}.
\end{equation}%
Then, (\ref{C0l0l}) follows from the identity
\begin{equation}
\sum_{j=0}^{A}\frac{r}{r+2j}{\binom{r+2j}{j}}{\binom{2(A-j)}{A-j}}={\binom{%
2A+r}{A},}  \label{fussid}
\end{equation}%
with $A=(n-m)l/2$ and $r=ml$. One way to verify Eq. (\ref{fussid}) is by
looking at the generating functions of the two sides. The left-hand side is
in the form of convolution $\sum_{j=0}^{A}a_{j}b_{A-j}$ with $a_{j}$ being
the Fuss-Catalan (or Raney) sequence, $\frac{r}{r+2j}{\binom{r+2j}{j}}$, and
$b_{j}$ the central binomial coefficients ${\binom{2j}{j}}$. The generating
function of the Fuss-Catalan numbers can be read off as $\sum
a_{j}z^{j}=\left( \frac{1-\sqrt{1-4z}}{2z}\right) ^{r}$ from (2.5.16) of
\cite{genfun}, while some more detailed analysis can be found in section 7.5
of \cite{concrete}. (The appearance of Fuss-Catalan numbers may seem
surprising in the context of nonlinear dynamics of PDEs, but they are
ubiquitous in random matrix and random tensor theory \cite%
{fc1,raney,fc2,fc3,fc4}.) The generating function of the central binomial
coefficients can be read off from (2.5.15) of \cite{genfun} as $\sum
b_{j}z^{j}=\frac{1}{\sqrt{1-4z}}$. The generating function of the
convolution $\sum_{j=0}^{A}a_{j}b_{A-j}$ is a product of the generating
functions of $a_{j}$ and $b_{j}$ and thus agrees with the generating
function of the right-hand side of Eq. (\ref{fussid}) that can be again read
off from (2.5.15) of \cite{genfun}. Thus, Eqs. (\ref{fussid}) and (\ref%
{C0l0l}) are valid.


\section{Vanishing of the coefficients $C_{2011}$ corresponding to (\protect
\ref{eq:stationary_solutions_family_1})}

\label{app:vanish}

We aim to prove that, starting with only two modes $%
(n_{0},m_{0})=(2i+1,-2j-1)$ and $(n_{1},m_{1})=(2k,0)$ with $j\leq i<k$ and $%
2(i-j)<k$, the energy does not flow in the resonant system (\ref%
{eq:Resonant_Eq_resonant}). As explained in section~\ref{sec:trunc}, in the
subsequent evolution only those modes get excited that lie on the straight
line connecting $(n_{0},m_{0})$ and $(n_{1},m_{1})$ in Fig.~\ref%
{fig:Diagram_nm}, yielding a simpler effective resonant system (\ref%
{eq:Resonant_Eq_resonant_nmkl}), where the two initial modes are labelled as
modes $0$ and $1$. Proving that there is no energy transfer out of these two
modes, as per discussion of section~\ref{sec:breath}, amounts to proving
that the interaction coefficient $C_{2011}$ in (\ref%
{eq:Resonant_Eq_resonant_nmkl}) vanishes. In terms of the original mode
tower of Fig.~\ref{fig:Diagram_nm}, this coefficient describes a quartic
interaction of mode $(2i+1,-2j-1)$, two copies of mode $(2k,0)$ and one mode
$(4k-2i-1,2j+1)$, and hence Eq. (\ref{CLaguerre}) makes it proportional to
\begin{equation}
\int_{0}^{\infty }d\rho \,e^{-2\rho }\,L_{i+j+1}^{-2j-1}(\rho
)\,L_{2k-i-j-1}^{2j+1}(\rho )\,(L_{k}^{0}(\rho ))^{2}.  \label{Cvanishdef}
\end{equation}%
We will now prove that this expression vanishes, confirming the
no-energy-transfer result of section~\ref{sec:breath}. (We ignore overall
numerical prefactors below, as they cannot affect a proof of the fact that
the quantity in question vanishes.)

Substituting the expression for the Laguerre polynomials in terms of their
generating function
\begin{equation}
L_{n}^{\alpha }(x)=\frac{1}{n!}\,\partial _{t}^{n}\frac{e^{-\frac{t}{1-t}x}}{%
(1-t)^{\alpha +1}}\Bigg|_{t=0}
\end{equation}%
and performing the integral over $\rho $ leaves an expression proportional
to
\begin{equation}
\partial _{t}^{i+j+1}\partial _{s}^{2k-i-j-1}\partial _{u}^{k}\partial
_{v}^{k}\left( \frac{1-t}{1-s}\right) ^{\hspace{-1mm}2j+1}\hspace{-3mm}%
\left( 2-t-s-u-v+tsu+tsv+tuv+suv-2tsuv\right) ^{-1}\Bigg|_{t,s,u,v=0}.
\label{Cdrv}
\end{equation}%
While we anticipate that an economical proof that this expression vanishes
may exist, for example based on complex-plane arguments, constructing this
proof appears challenging and we present below a brute force proof based on
an analytic evaluation of (\ref{Cdrv}) and book-keeping cancellations among
various terms that arise. (For any particular values of $i$, $j$ and $k$, it
is of course straightforward to evaluate the above expression and check that
it vanishes.)

The $u$ and $v$ derivatives in (\ref{Cdrv}) only act on the last factor and
can be evaluated explicitly. For the $v$-derivative, ignoring the overall
prefactor, we get
\begin{equation}
\partial _{v}^{k}\left( 2-t-s-u-v+tsu+tsv+tuv+suv-2tsuv\right) ^{-1}\Bigg|%
_{v=0}=\frac{(1-ts-tu-su+2tsu)^{k}}{(2-t-s-u+tsu)^{k+1}}.
\end{equation}%
Then, acting on this expression with $k$ $u$-derivatives and binomially
distributing them between the numerator and denominator (setting $u=0$
thereafter) produces a collection of terms proportional to
\begin{equation}
\frac{(1-ts)^{2p}\,(2ts-t-s)^{k-p}}{(2-t-s)^{k+p+1}},
\end{equation}%
where $p$ ranges from 0 to $k$. It turns out that these terms individually
give vanishing contributions to (\ref{Cdrv}), which we now proceed to prove
by showing that
\begin{equation}
\partial _{t}^{i+j+1}\partial _{s}^{2k-i-j-1}\left( \frac{1-t}{1-s}\right)
^{2j+1}\frac{(1-ts)^{2p}\,(2ts-t-s)^{k-p}}{(2-t-s)^{k+p+1}}\Bigg|_{t=s=0}
\label{drvts}
\end{equation}%
vanishes, which implies that (\ref{Cdrv}) vanishes.

We start out by expanding $(1-ts)^{2p}$ as
\begin{equation}
(1-ts)^{2p}=2^{-2p}(2-t-s-(2ts-t-s))^{2p}=2^{-2p}\sum_{q=-p}^{p}{\binom{2p}{%
p+q}}(-1)^{p+q}(2-t-s)^{p-q}(2ts-t-s)^{p+q}.
\end{equation}%
It turns out that each pair of terms with $q=r$ and $q=-r$ in this sum gives
contributions that cancel each other in Eq. (\ref{drvts}); in other words,
we must prove that
\begin{equation}
\partial _{t}^{i+j+1}\partial _{s}^{2k-i-j-1}\left( \frac{1-t}{1-s}\right)
^{2j+1}\left( \frac{(2ts-t-s)^{r}}{(2-t-s)^{r+1}}+\frac{(2ts-t-s)^{2k-r}}{%
(2-t-s)^{2k-r+1}}\right) \Bigg|_{t=s=0}=0  \label{pairwise}
\end{equation}%
for any $r$ between 0 and $k$, and this will yield Eqs. (\ref{drvts}), (\ref%
{Cdrv}) and hence (\ref{Cvanishdef}).

Consider the generating function for the entries featured in (\ref{pairwise}%
), defined by
\begin{equation}
\mathcal{F}_{ijk}(z)\equiv \sum_{r=0}^{\infty }z^{r}\,\,\partial
_{t}^{i+j+1}\partial _{s}^{2k-i-j-1}\left( \frac{1-t}{1-s}\right) ^{2j+1}%
\frac{(2ts-t-s)^{r}}{(2-t-s)^{r+1}}\Bigg|_{t=s=0}.
\end{equation}%
We will prove that $\mathcal{F}_{ijk}$ is a polynomial in $z$ of degree $2k$
satisfying
\begin{equation}
z^{2k}\mathcal{F}_{ijk}(1/z)=-\mathcal{F}_{ijk}(z).  \label{invpoly}
\end{equation}%
As a result, the coefficients of $z^{r}$ and $z^{2k-r}$ in this polynomial
have the same magnitude and opposite signs, which hence secures the validity
of Eq. (\ref{pairwise}). To prove this statement, we evaluate the sum over $%
r $ as a geometric series to obtain
\begin{align}
\mathcal{F}_{ijk}(z)& =\partial _{t}^{i+j+1}\partial _{s}^{2k-i-j-1}\left(
\frac{1-t}{1-s}\right) ^{2j+1}\left( 2-t-s+tz+sz-2tsz\right) ^{-1}\Bigg|%
_{t=s=0}  \notag \\
& =-(i+j+1)!\,(z+1)^{2j+1}\,\partial _{s}^{2k-i-j-1}\frac{(1-z+2sz)^{i-j}}{%
(2-s+sz)^{i+j+2}}\Bigg|_{s=0},  \notag
\end{align}%
where, to arrive at the last representation, one distributes $\partial
_{t}^{i+j+1}$ binomially between $(1-t)^{2j+1}$ and $\left(
2-t-s+tz+sz-2tsz\right) ^{-1}$, performs the elementary differentiation,
sets $t=0$, and thereafter resums the resulting binomial expression.
Finally, distributing $\partial _{s}^{2k-i-j-1}$ binomially between $%
(1-z+2sz)^{i-j}$ and $(2-s+sz)^{-(i+j+2)}$ in the last line, differentiating
and setting $s=0$ results in a collection of terms of the form of $%
z^{q}(z+1)^{2j+1}(z-1)^{2k-2j-2q-1}$ with $q$ ranging from $0$ to $i-j$.
However, each such term is a polynomial of degree at most $2k$, manifestly
satisfying Eq. (\ref{invpoly}). Being made entirely out of such terms, $%
\mathcal{F}_{ijk}$ satisfies (\ref{invpoly}), which ensures the validity of
Eq. (\ref{pairwise}), hence (\ref{drvts}) and (\ref{Cdrv}) vanish, as well
as the interaction coefficient (\ref{Cvanishdef}).

\twocolumngrid

\end{document}